\documentclass[amsmath,amssymb,noshowpacs,aip,pop,
preprint,
citeautoscript  
]{revtex4-1}
\usepackage[mathscr]{eucal}
\usepackage{amssymb}
\usepackage{amsmath}
\usepackage{dcolumn}
\usepackage{bm}
\usepackage{graphicx}
\usepackage{epstopdf}
\usepackage{subfigure}
\usepackage{color}
\usepackage{xcolor}
\usepackage{setspace}
\usepackage[percent]{overpic}
\usepackage{booktabs}
\usepackage{lineno}
\usepackage{amsthm}

\usepackage{geometry}
\newtheoremstyle{query}%
{}{}
{\color{red}}
{}
{\sffamily\bfseries}{:}{12pt}
{}
\theoremstyle{query}

\UseRawInputEncoding
\begin{document}

\title{Gyrokinetic simulations of electric current generation in ion temperature gradient driven turbulence}

\author{Xiang Chen}
\affiliation{School of Nuclear Science and Technology, University of Science and Technology of China, Hefei 230026, People's Republic of China}

\author {Zhixin Lu}
\email {zhixin.lu@ipp.mpg.de}   
\affiliation{Max Planck Institut f\"ur Plasmaphysik, 85748 Garching, Germany}

\author{Huishan Cai}
\email {hscai@mail.ustc.edu.cn}
\affiliation{School of Nuclear Science and Technology, University of Science and Technology of China, Hefei 230026, People's Republic of China}

\author{Lei Ye}
\affiliation{Institute of Plasma Physics, Chinese Academy of Sciences, Hefei 230031, People's Republic of China}

\author{Yang Chen}
\affiliation{Center for Integrated Plasma Studies, University of Colorado at Boulder, Boulder, CO 80309, United States}

\author{Baofeng Gao}
\affiliation{School of Nuclear Science and Technology, University of Science and Technology of China, Hefei 230026, People's Republic of China}

\date{\today}

\begin{abstract}
Gyrokinetic simulations in the collisionless limit demonstrate the physical mechanisms and the amplitude of the current driven by turbulence.
Simulation results show the spatio-temporal variation of the turbulence driven current and its connection to the divergence of the Reynolds stress and the turbulence acceleration. 
Fine structures (a few ion Larmor radii) of the turbulence induced current are observed near the rational surfaces with the arbitrary wavelength solver of the quasi-neutrality equation. The divergence of the Reynolds stress plays a major role in the generation of these fine structures. 
The so-called ``spontaneous'' current is featured with large local magnitude near the rational surfaces.

\end{abstract}

\maketitle

\section{Introduction}
\label{introduction}

Non-inductive current is essential to magnetic fusion experiments, such as neutral beam injection current drive, lower hybrid
current drive, electron cyclotron current drive and bootstrap current. Compared with the external current drive, the bootstrap current
is driven by the radial density and temperature gradient, and thus is more economical\cite{bickerton1971,peeters2000}. 
However, bootstrap current brings in Magnetohydrodynamic (MHD) instabilities such as neoclassical tearing modes (NTMs). Besides the bootstrap current, the turbulence driven current is also a kind of intrinsic current in tokamak plasma. 
Turbulence is widely observed in tokamak operations.
It affects the electron parallel momentum transport through different mechanisms such as the electron momentum flux and the electron-ion momentum exchange. The spontaneous variation of electron momentum provides a current source which corrugates the current density profile and has significant impacts on the generation of the plasma self-driven mean current\cite{mcdevitt2017,wang2019}. The current induced by turbulence affects the MHD instabilities which can provide a way to study the interaction between the turbulence and the MHD instabilities. For example, this spontaneous current can enhance or reduce the onset threshold of NTMs depending on the direction of the turbulence intensity gradient\cite{cai2018}. 

The current driven by turbulence has been discussed in detail using the analytical and numerical approaches\cite{ITOH1988,hinton2004,garbet2014,yi2016,mcdevitt2017,he2018,wang2019}.  
It has been shown by S. -I. Itoh and K. Itoh firstly with a slab model \cite{ITOH1988}. Drift-wave fluctuations drive a net current when   
$\left\langle k_{\parallel}\right\rangle$ is finite, where $\left\langle k_{\parallel}\right\rangle$ is the averaged parallel wave number over the spectrum .
Hinton proposed that turbulence drives current by providing a dynamo electromotive force (EMF)\cite{hinton2004}. EMF is mainly produced by two mechanisms. One is the divergence of the electron parallel momentum flux which is produced by the magnetic flutter, and the other is the beating of the parallel electric field fluctuations with the electron density fluctuations\cite{hinton2004}.
In the electrostatic turbulence, three mechanisms of the turbulence driven current have been analyzed by McDevitt \cite{mcdevitt2017}, namely, 
(1) the electron residual stress, which causes a current redistribution,
(2) the turbulent acceleration, which is the momentum exchange between ions and electrons and causes a net current, 
(3) the turbulence induced resonant electron scattering, which leads to the equilibrium between the trapped and passing electrons\cite{McDevitt2013}, similar to the mechanism of the bootstrap current\cite{bickerton1971,peeters2000}. 
Electromagnetic simulations show that the dynamo current density has spike structures near the rational surface\cite{hinton2004}. The amplitude of the spike structures is about 150$\%$ of the Ohmic current density， and the total current is about 1$\%$ of Ohmic current for DIII-D parameters. 
A direct calculation of electron and ion current generation by ITG and CTEM turbulence was reported in early global GK simulations\cite{wang2012}. The connection between the electron flow generation and the residual stress and the turbulence acceleration is discussed by Wang\cite{wang2019}.
The electrostatic studies by Yi \cite{yi2016} show that the net current induced by electron temperature gradient turbulence is approximately $20\%$ of the local bootstrap current density.
Most previous simulations of the current flow generation in tokamak plasmas are based on the long wave-length approximation in the quasi-neutrality equation, and the fine structures of the turbulence driven current have not been thoroughly investigated. In this work, the fine structures of the mode \cite{waltz2006,dominski2017} near the rational surfaces are observed using the gyrokinetic field solver\cite{chen2007,chen2010}.  These spike structures of the current are investigated with the gyrokinetic simulation in electrostatic ITG turbulence.

The gyrokinetic code GEM\cite{chen2003,chen2007} has been used to study the ITG turbulence induced current in the collisionless limit. 
The electron momentum flux and turbulent electron-ion momentum exchange have been analyzed for identifying the current driven mechanisms, to draw the connection with the theoretical analyses \cite{mcdevitt2017}. 
The remainder of this paper is organized as follows. 
In Sec.~\ref{model}, numerical tools and the parameters are described in detail.
In Sec.~\ref{mechanisms and results}, physical mechanisms of the turbulence driven current and the results of multiple mode simulations are discussed.
Finally, the paper is concluded in Sec.~\ref{conclusions}.

\section{Simulation model and employed parameters}
\label{model}
\subsection{Simulation model}

This work is carried out by using the simulation code GEM\cite{chen2003,chen2007}. GEM is a $\delta f$ gyrokinetic Particle-in-cell code for the study of low-frequency phenomenon such as the micro-turbulence and energetic particle driven Alf\'ven modes in tokamak plasmas. It solves the gyrokinetic Vlasov-Maxwell equations with gyrokinetic ions and drift-kinetic electrons, and neglects the parallel magnetic perturbations. 
In the $\delta f$-PIC method, the perturbed distribution is sampled by the marker particles in the 5D phase space. 
GEM evolves the orbit of marker particles in the Lagrangian frame and calculates the weight $w=\delta f/f$ at each time step.  
The charge and the current density are calculated using the weight and the coordinates of marker particles, and are used to solve the quasi-neutrality and Amp\`ere equations. In the quasi-neutrality equation $-q_in_p=q_i\overline{\delta n}_i - e\delta n_e$, the electric potential $\phi$ is calculated with arbitrary wavelength solver $n_p\propto \sum_{k_\perp}{e^{ik_{\perp}\cdot x}\phi_{k_{\perp}}\left[ 1-\Gamma_0(b)\right]}$\cite{chen2007}, where $b=k_{\perp}^2v_T^2/\Omega_i^2$ denotes the ion finite Larmor radius effects, $\Gamma$ is the Gamma function , $\phi_{k_{\perp}}$ represents the perpendicular Fourier components of $\phi$, $q_i$ is the ion charge, $n_p$ is the  polarization density, $\overline{\delta n}_i$ is the perturbed density of guiding centers  and $\delta n_e$ is the perturbed density of electron guiding centers. 
The perturbed electron distribution function $\delta f_e$ is split into the adiabatic and non-adiabatic parts \cite{lee2001,chen2007} for treating the fast parallel dynamics of electrons, 

\begin{align}
    \delta f_e=\epsilon_g\frac{e\delta\phi}{T_e}f_{0e}+\delta h_e\;\;,
\end{align}
where $\epsilon_g$ is an adjustable parameter, $ef_{0e}\epsilon_g \delta\phi/T_e$  and $\delta h_e$ are the adiabatic part and  non-adiabatic part of $\delta f_e$. GEM uses the parallel canonical momentum as a velocity coordinate
$p_{\parallel}=v_{\parallel}+q/m\left\langle A_{\parallel}\right\rangle$,
where $v_{\parallel}$ is the parallel guiding center velocity, $A_{\parallel}$ is the fluctuating parallel vector potential, 
$\left\langle \cdots \right\rangle$ represents the gyro-average. 
The field-aligned coordinates $\left(x,y,z\right)$ are used to push the marker particles,
\begin{equation}
x=r-r_c \;\;,y=\frac{r_c}{q_c}\left(\int^{\theta}_0\hat{q}\left(r,\theta^\prime\right)d\theta^{\prime}-\zeta\right) \;\;,z=q_cR_0\theta
\end{equation}
where, $\hat{q}\left(r,\theta\right)=\boldsymbol{B}\cdot\nabla\zeta/\boldsymbol{B}\cdot\nabla\theta$,
$\left(r,\theta,\zeta \right)$ denote the minor radius, the poloidal angle and the toroidal angle respectively, $r_c$ and $q_c$ are the minor radius
and safety factor at the center of the simulation domain.

\subsection{Parameters setting and benchmark}
In the following simulations, the D\uppercase\expandafter{\romannumeral3}-D Cyclone Base Case (CBC)\cite{dimits2000} parameters are adopted. 
The parameters are the same as those in a previous benchmark work\cite{gorler2016}.
The deuterium ($m_i/m_p=2$) is the only ion species and the kinetic electrons with an electron mass of $m_e/m_p=1/1837$ are considered. The concentric circular magnetic equilibrium with inverse aspect ratio $a/R_0=0.36$ is adopted for simplicity. The safety factor profile is\cite{gorler2016} 
$q(r)=2.52(r/a)^2-0.16(r/a)+0.86$, where $a$ and $R_0$ denote the minor and major radius, $r$ is the local minor radius and the magnetic shear is defined as 
$\hat{s}=d\ln q/d\ln r$. 
The equilibrium density and temperature profiles are indicated by A(r) and the normalized logarithmic gradients are
defined as $L_{\mathrm{ref}}/L_{A}$ 

\begin{flalign}
     \frac{A(r)}{A(r_0)}  =\exp\left[-\kappa_Aw_A\frac{a}{L_{\mathrm{ref}}}\tanh\left(\frac{r-r_0}{w_Aa}\right)\right]\;\;,   \\ 
     \frac{L_{\mathrm{ref}}}{L_{A}} =-L_{\mathrm{ref}}\frac{d\ln A(r)}{dr}=\kappa_A\cosh^{-2}\left(\frac{r-r_0}{w_Aa}\right)\;\;,
\end{flalign}
where $L_{\mathrm{ref}}$ is the macroscopic reference length which is equal to $R_0$ in the following, and 
$L_A=-\left(d\ln A(r)/dr\right)^{-1}$.
At the reference radius $r_0$ $(r_0/a=0.5)$, density and temperature 
profiles have a peaked normalized logarithmic gradient with the characteristic width $w_A$ and the maximum amplitude $\kappa_A$.
Other reference parameters are listed in Table \ref{paramt_of_CBC}. 
The dimensionless parameter $\rho^{\ast}=\rho_s/a=(c_s/\Omega_i)/{a}$ is approximately 1/180, where $c_s=\sqrt{T_{ref}/m_{i}}$ is the ion sound speed and $\Omega_i=eB_{0}/m_i$ is the ion cyclotron frequency at the magnetic axis and $B_{\mathrm{0}}$ is the toroidal magnetic field on axis. 
$\beta_e$ is calculated according to $\beta_e= n_{\mathrm{e}}T_{\mathrm{e}}/(B^2_{\mathrm{0}}/2\mu_0)$ and scanned by varying $n_{\mathrm{e}}$, where $n_{\mathrm{e}}$ and $T_{\mathrm{e}}$ are the elctron density and temperature taken at $r_0$ respectively. The $\beta_e$ scan of GEM with fixed toroidal mode number $(n=19)$ is performed in Ref. \onlinecite{ye2020}. The linear frequency and growth rate of the ITG modes and KBMs are very close to the GENE results in the range of $\beta_e<2.5\%$. In the following studies of the current generation, we focus on the ITG turbulence in the electrostatic limit ($\beta_e=0.1\%$ by default except clarified) and the electromagnetic effects will be studied in future work.

\begin{table}[htbp]
\caption{Parameters of Cyclone Base Case, $R_0$ and $B_0$ denote major radius and magnetic field on the magnetic axis, $\nu_{coll}=0$ represents the 
collisionless simulation in this work.}
\centering
\begin{tabular*}{100mm}{@{\extracolsep{\fill}}c c }
\hline
$R_0$ & 1.67m \\
$B_0$ & 2.0T \\
$T_i(r_0)=T_e(r_0)=T_{ref}$ & 2.14 keV \\
$\kappa_{Ti}=\kappa_{Te}$ & 6.96 \\
$w_{Ti}=w_{Te}$ & 0.3 \\

$n_i(r_0)=n_e(r_0)=n_{ref}$ & $4.66\times10^{19}m^{-3}$ \\
$\kappa_{ni}=\kappa_{ne}$ & 2.23 \\
$w_{ni}=w_{ne}$ & 0.3 \\
$\nu_{coll}$ & 0 \\
$q(r_0)$ & 1.4\\
$\rho^*=\rho_s/a$ &$5.56\times 10^{-3}$ \\
\hline
\end{tabular*}
\label{paramt_of_CBC}
\end{table}

\section{Turbulence driven current mechanisms and simulation results}
\label{mechanisms and results}
\subsection{Turbulence driven current mechanisms}
Turbulence has effects on the electron parallel momentum transport and drives the so-called ``spontaneous'' current\cite{yi2016}.
In the collisionless and electrostatic limit, the electron parallel momentum equation\cite{abiteboul2011,mcdevitt2017} can be written as

\begin{align}
\label{Current_derivation}
\frac{\partial}{\partial t}j_{\parallel e} = -\frac{q_e}{m_e}\nabla\cdot \Pi_{\parallel e} -\frac{q_e}{m_e}M_{\parallel e} \;\;,
\end{align}
where $j_{\parallel e} = -e\int d^3vv_{\parallel}\delta f_e$ is the parallel electron current produced by turbulence, $\Pi_{\parallel e}$ is the electron parallel momentum flux and $M_{\parallel e}$ denotes the turbulence acceleration. 
Equation (\ref{Current_derivation}) demonstrates that the variation of turbulence driven current is produced by the divergence of Reynolds stress and the turbulence acceleration.
In the electrostatic limit, the electron parallel momentum flux $\Pi_{\parallel e}$ can be expressed as

\begin{align}
\label{eq:pi_definition}
    \Pi_{\parallel e}=m_e\Re\left\langle {\overline{{\int}{d^{3}v v_{\parallel}\delta f_e
  \delta\boldsymbol{v}_{EB} \cdot \boldsymbol{\hat{r}}}}} \right\rangle_{FS}  \;\;,
\end{align}
where $\left( \overline{\cdots} \right)$ is the temporal average and $\left\langle{\cdots}\right\rangle_{FS}= \iint{(\cdots)\mathcal{J}d\theta d\zeta }/\iint{\mathcal{J}d\theta d\zeta}$ represents the flux surface average. $\delta f_e$ is the perturbed electron distribution function, $v_\parallel$ is the electron parallel velocity, $\delta\boldsymbol{v}_{EB}$ denotes the $\boldsymbol{E\times B}$ drift velocity and $\boldsymbol{\hat{r}}$ is the unit vector in radial direction.
The second term of the right hand side of Eq.~(\ref{Current_derivation}), the electron-ion momentum exchange $M_{\parallel e}$, can produce a net current by modifying the electron momentum. In the electrostatic limit, it is given by

\begin{align}
    M_{\parallel e}= q_e\left\langle {\overline{{\int}{d^{3}v \delta f_e
  \hat{\boldsymbol{b}} \cdot \nabla\delta\phi }}} \right\rangle_{FS}  \;\; ,
\end{align}
where $\hat{\boldsymbol{b}}=\boldsymbol{B}/|B|$ and $\delta \phi$ is the perturbed electric potential. 

\subsection{Simulation results and analyses}
In the following simulations, a simplified model with single ITG mode $(n=20)$ and zonal mode $(n=0)$ are used, where the simulation domain is $1/20$th of torus. The full dynamics of gyrokinetic ions and drift-kinetic electrons are included with realistic electron-ion mass ratio. The grid numbers for $(x,y,z)$ are set as $n_x=256$, $n_y=64$ and $n_z=64$ respectively.  Convergence studies have been done by comparing cases with $n_x=256,384,512$ and good convergence has been observed when $n_x\ge256$.  The time steps considered is $\Delta t=1/\Omega_p$, where $\Omega_p$ is the proton cyclotron frequency $\Omega_p=eB_0/m_p$. The marker number per cell is $32$ for both ions and electrons.
\subsubsection{Time evolution of the physical quantities}

To understand the mechanism of turbulence driven current, 
it is necessary to consider the time evolution of the physical quantities in Eq.~(\ref{Current_derivation}). 
Since $\delta F_e \propto \delta \phi $ and $\delta \boldsymbol{v_{EB}}=\hat{\boldsymbol{b}}\times\nabla\delta\phi/B\propto\delta \phi$, it is expected that  
$\{I,\ q_e/m_e\nabla \cdot \Pi_{\parallel e},\ q_e/m_e M_{\parallel e}\}$ are proportional to $\delta\phi^2$ and grow up at about twice ITG growth rate in the linear stage before the saturation in nonlinear regimes. Here, $I$ denotes the flux surface averaged turbulence intensity $I=\left\langle\delta\phi^2\right\rangle_\mathrm{FS}$.
In Fig. \ref{Physical_quantities_with_t}, physical quantities evolve with time near the rational surface ($q(r)=1.3$)  where the turbulence signal is strong.
The blue line represents $\partial \delta j_{\parallel e}/\partial t$, the yellow line is the turbulent intensity $I$, the green line denotes $q_e/m_e\nabla \cdot \Pi_{\parallel e}$, the red line is $q_e/m_e M_{\parallel e}$ and the black lines are their exponential fitting in the linear stage. They grow exponentially at almost the same growth rate in linear stage,  and reach saturation after $t=18R_0/c_s$.

\subsubsection{Spatial structures of $\Pi_{\parallel e}$ and $M_{\parallel e}$ in the linear stage} 
In this section, the radial profile of $\Pi_{\parallel e}$ and $M_{\parallel e}$ are discussed. With electrostatic approximation, the electron parallel momentum flux can be written as\cite{gurcan2007,garbet2014,mcdevitt2017} 

\begin{align} 
\label{Pi_comp}
\Pi_{\parallel e}=-\chi_{\phi}\frac{\partial \bar{v}_{\parallel e}}{\partial r}+V\bar{v}_{\parallel e}+\pi_{\parallel e}\;\; ,
\end{align}
where $\bar{v}_{\parallel e}$ is the mean electron parallel flow velocity. 
The first term on the right hand side is the electron viscosity with $\chi_{\phi}$ being the turbulent viscosity coefficient, the second term is a pinch of electron momentum with $V$ being the coefficient of the electron momentum pinch and the last term is regarded as the electron residual stress which is independent of parallel flow and its gradient.
In our particle simulation with Maxwellian electron distribution as the initial condition, the first and second terms do not contribute to the $v_{\parallel e}$ generation at the beginning, and the residual stress $\pi_{\parallel e}$ is the key to initial the $v_{\parallel e}$ generation due to turbulence. The turbulence intensity gradient induced residual stress has been studied in the ion rotation problems \cite{gurcan2010,lu2015}, and we analyze the variable $\partial I/\partial r$ to identify its contribution to the generation of electron residual stress ($\Pi_{\parallel e} \sim \pi_{\parallel e}$).

For the cyclone base case studied here, there is no up-down asymmetry in the equilibrium \cite{Camenen2009}, and there is no equilibrium flow shear effects.\cite{wang2009} During the linear stage, residual stress is mainly
produced by symmetry breaking mechanisms such as the intensity gradient mechanism and the poloidal tilt of the global mode structure. Formally, the residual stress can be described as $\pi_{\parallel e}=\alpha \hat{s} \partial I/\partial r+\pi^{tilt}_{\parallel e}$\cite{gurcan2010,Camenen2011,lu2015pop,lu2017}, where $\alpha$ is a coefficient determined by the equilibrium parameters and the fluctuation properties, and  $\pi^{tilt}_{\parallel e}$ is the residual stress induced by the profile shearing effects,  related to the ``tilting angle'' of the two dimensional mode structure. The gradient of turbulence intensity  breaks the symmetry property of the parallel mode structure and leads to the net residual stress\cite{gurcan2010,lu2017}. 
Figure \ref{figure9}(a) and Fig. \ref{figure9}(b) are the structures of turbulence intensity and its radial gradient in linear stage. GEM solves the quasi-neutrality equation with arbitrary wavelength solver\cite{chen2007,chen2010}, and the fine structures near the rational surface are captured. In addition, drift kinetic electrons with realistic mass ratio are included in the simulations. Thus the resonance between the ITG mode and the fast moving electrons near the rational surface is included in the simulations, which can be an important ingredient of the fine structures generation.
Figure \ref{figure9}(b) and Fig. \ref{figure9}(c) show that the turbulence intensity gradient and electron residual stress have odd parity and increase sharply near the rational surface. The blue line of Fig. \ref{figure9}(c) and Fig. \ref{figure9}(d) are the radial profile of the Reynolds stress $\Pi_{\parallel e}$ and its divergence $a\nabla\cdot\Pi_{\parallel e}$.  Along the radial direction, $\Pi_{\parallel e}$ increases sharply near the rational surface and decreases more slowly between two rational surface. 
Near the rational surface, the magnitude of $a\nabla \cdot\Pi_{\parallel e}$ is about $170$ times of the variation of $\Pi_{\parallel e}$.
Since $a\nabla\cdot\Pi_{\parallel e}\sim a/L_{\Pi}\Pi_{\parallel e}$, it is obtained that $\Pi_{\parallel e}/(a\nabla\cdot\Pi_{\parallel e})\sim L_\Pi/a\sim 1/170$, where $L_\Pi$ denotes the characteristic length of $\Pi_{\parallel e}$ near the rational surface. The characteristic length of $\Pi_{\parallel e}$ ($L_{\Pi}$) is the scale of the ion Larmor radius ($\rho_i$) near the rational surface. Hence $\nabla\cdot\Pi_{\parallel e}$ has fine structures in the vicinity of the rational surface. The width of the fine structures is close to $\rho_i$, which is consistent with the theoretical analyses of previous work \cite{hinton2004}. Besides these, the 
fine structures of $\nabla\cdot\Pi_{\parallel e}$ broaden slightly with time as Fig. \ref{figure2} shows. 

Residual stress redistributes the profile of electron parallel momentum but does not change the total electron parallel momentum. The turbulent acceleration term describes the turbulence induced momentum exchange between ions and electrons\cite{mcdevitt2017} and acts as a local source or sink\cite{wanglu2013}.  Therefore, its effects on turbulence driven current are meaningful and potentially important. In Fig. \ref{figure9}(d), the blue line is the divergence of Reynolds stress ($\nabla\cdot\Pi_{\parallel e}$) and the red line denotes the turbulence acceleration ($M_{\parallel e}$). $aM_{\parallel e}$ is much smaller than $a\nabla\cdot\Pi_{\parallel e}$ which shows that the ITG turbulence induced current is mainly produced by the divergence of electron Reynolds stress in linear stage.

\subsubsection{Analyses of the current generation mechanism due to $\nabla\cdot\Pi_{\parallel e}$  and $M_{\parallel e}$}
The dominant mechanism of the turbulence driven current is identified by analyzing the contributions from $\nabla\cdot\Pi_{\parallel e}$  and $M_{\parallel e}$.  
Figures ~\ref{figure3}$\left(a\right)$, \ref{figure3}$\left(b\right)$ and \ref{figure3}$\left(c\right)$ show the linear radial structures of the divergence of electron parallel momentum flux $\nabla\cdot\Pi_{\parallel e}$, the electron-ion momentum exchange $M_{\parallel e}$ and the time variation of turbulence driven current $\partial j_{\parallel e}/\partial t$ in $r/a\in\left[0.3,0.65\right]$. 
The fine structures of $\nabla\cdot\Pi_{\parallel e}$, $M_{\parallel e}$ and $\partial j_{\parallel e}/\partial t$ centered around the rational surface are evident. The radial scale of the fine structures is about $\rho_i$ which is discussed above. 
For $r/a<0.47$, $M_{\parallel e}$ is positive and for $r/a>0.47$, $M_{\parallel e}$ is mainly negative except that it is near the rational surface. 
The divergence of Reynolds stress has a large value on the rational surface and its contribution to $\partial j_{\|e}/\partial t$ is much larger than that from $M_{\parallel e}$. 
Figure \ref{figure3} $\left(d\right)$ shows the quantitative comparison of the left and right hand side terms of Eq.~(\ref{Current_derivation}).
Near the rational surface, $-q_e/m_e\left(\nabla\cdot\Pi_{\parallel e}+M_{\parallel e}\right)$ (blue line) and $\partial j_{\parallel e}/\partial t$ (red line) have similar spike structures and the same magnitude in linear regime. The ratio between $-q_e/m_e\left(\nabla\cdot\Pi_{\parallel e}+M_{\parallel e}\right)$ and $\partial j_{\parallel e}/\partial t$ $\left(at\ r/a=0.43\right)$ fluctuates and increases gradually around 1 as Fig. \ref{Ratio_pJpt_M_Pi} shows. These indicate that the time variation of the intrinsic current is mainly produced by $\nabla\cdot\Pi_{\parallel e}$ and $M_{\parallel e}$.

Besides the linear analyses, nonlinear behaviors are studied in the following, by  calculating the correlations among different variables. 
The correlation coefficient between two profiles at the same time is expressed as follows:
 
\begin{align}
  C[A,B]=\frac{\sum_i[A(r_i)-\bar{A}][B(r_i)-\bar{B}]}{\sqrt{\sum_i[A(r_i)-\bar{A}]^2\sum_i[B(r_i)-\bar{B}]^2}}\;\;.
\end{align}
$\bar{A}$ and $\bar{B}$ are the mean value over $r$ where the fluctuations are more intensive. The closer the absolute value of the coefficient is 
to 1, the stronger the correlation between A and B is. The coefficient less than 0 indicates that A and B are negatively correlated.
Figure~\ref{correlation_pJpt_M_Pi} shows the correlations among the time variation of the turbulence induced current ($\partial j_{\parallel e}/\partial t$), the divergence of the Reynolds stress ($\nabla\cdot\Pi_{\parallel e}$), the turbulence acceleration ($M_{\parallel e}$), the turbulence intensity gradient ($\partial I/\partial r$) and the second derivative of the turbulence intensity ($\partial^2 I/\partial r^2$). For the current generation problem in this work, significant correlation between the $\nabla\cdot\Pi_{\parallel e}$ and $\partial j_{\parallel e}/\partial t$ has been observed as shown by the blue line in Fig.~\ref{correlation_pJpt_M_Pi}. 
From Eq.~(\ref{Current_derivation}), in collisionless limit, turbulence induced current is contributed by $\nabla\cdot\Pi_{\parallel e}$ and $M_{\parallel e}$. Due to the fast variation of $\Pi_{\parallel e}$ $(L_{\Pi}\sim\rho_i)$ near the rational surface, the divergence of $\Pi_{\parallel e}$ is much greater than $M_{\parallel e}$. Hence, it is expected that $\partial j_{\parallel e}/\partial t$ is mainly produced by $\nabla\cdot\Pi_{\parallel e}$, which is consistent with the low correlation between $M_{\parallel e}$ and $\partial j_{\parallel e}/\partial t$ as indicated by the red line.
In previous work\cite{wang2009,gurcan2010,Camenen2011,lu2015,lu2017,hornsby2018global} of the ion toroidal intrinsic rotation of thermal ions, it is demonstrated that the  residual stress can be produced by the turbulence intensity gradient and the mode structure 'tilting' effect. In Fig.~\ref{correlation_pJpt_M_Pi}, from $t=7000t_u$ to $t=15	000t_u$, the correlation between $\Pi_{\parallel e}$ and $\partial I/\partial r$ (yellow line) indicates that a significant portion of $\Pi_{\parallel e}$ is due to the turbulent intensity gradient in linear stage. After the saturation of the ITG modes, the correlation $C[\Pi_{\parallel e},\partial I/\partial r]$ becomes lower. Because the turbulence intensity gradient decreases as nonlinear turbulence spreading occurs, other symmetry breaking mechanism become important, such as the symmetry breaking effects due to the zonal flow shear \cite{wang2009}.  
The strong correlation between $\partial j_{\parallel e}/\partial t$ and $\partial^2 I/\partial r^2$ (purple line) suggests that the curvature (second derivative) of the turbulence intensity significantly contributes to the generation of the spontaneous current. 
Note that even in the linear stage, the correlation varies along time. One possible reason is the time variation of the radial width of the fine structures in $\delta\phi$. In the early linear stage, the width of the fine structure is smaller than that in later stage and thus, the turbulent viscosity can mitigate the generation of the parallel momentum flux more significantly.

The generation of the electron parallel momentum flux $\Pi_{\parallel e}$ is related to the symmetry breaking of the mode structures\cite{gurcan2010,peeters_2011,lu2017}. 
With $\delta\boldsymbol{v_{EB}}=\hat{\boldsymbol{b}}\times\nabla\delta\phi/B$, $\Pi_{\parallel e}$ can be expressed as $\Pi_{\parallel e}\propto\left\langle k_{\parallel}\delta\phi^2_k\right\rangle$. $\left\langle k_{\parallel}\delta\phi^2_k\right\rangle$ denotes the spectrum weighted parallel wave vector, and is produced by the gradient of the turbulence intensity envelope\cite{gurcan2010}. 
Hence, to ignore the effects of the fine structures on $\left\langle k_{\parallel}\delta\phi^2_k\right\rangle$, the window average method are used. For a radial profile $G(r)$, $\left\langle G\right\rangle_{\mathrm{win}}$ represents its large scale structure computed by window average. The width of the windows are about the distance between the rational surface ($\sim 1/nq^\prime$). Figure~\ref{correlation_Pi_k_paral} indicates the correlation of the large scale structure between $\Pi_{\parallel e}$ and $\left\langle k_{\parallel}\delta\phi^2_k\right\rangle$ versus time. The strong correlation between $\left\langle\Pi_{\parallel e}\right\rangle_{\mathrm{win}}$ and $\left\langle k_{\parallel}\right\rangle_{\mathrm{win}}$ in linear stage shows that the averaged $k_{\parallel}$ symmetry breaking leads to the net electron momentum flux. In nonlinear stage, the effects of the symmetry breaking produced by the up-down asymmetry of the magnetic equilibrium\cite{Camenen2009}, profile shearing\cite{Camenen2011} and zonal flow shear\cite{wang2009} are also important. Hence the correlation between 
$\left\langle\Pi_{\parallel e}\right\rangle_{\mathrm{win}}$ and $\left\langle k_{\parallel}\right\rangle_{\mathrm{win}}$ decreases. 
In this work, we have closely followed the equations in the theoretical derivation \cite{mcdevitt2017} for diagnosis and more comprehensive analyses with the consideration of nonlinear dynamics such as the ZFs rely on the further development in that model in the future.

\subsubsection{Magnitude and scaling of current generation due to turbulence}

The simplified model included single ITG mode $(n=20)$ and zonal mode $(n=0)$ aims for demonstrating the underlying physics. However, this model is not accurate enough for predicting the amplitude of turbulence induced current. For example, the fine structures become less visible as more and more toroidal modes are included. \cite{waltz2006,dominski2017}  In this subsection, in order to understand the impact of turbulence induced current on the equilibrium current, multiple toroidal modes $n$ are simulated. To suppress the TEM instabilities of CBC, the multiple-$n$ simulations are carried out by using the lower electron temperature gradient $(\kappa_{Te}=2.23)$. Beside this, since we focus on the electrostatic ITG studies, a lower $\beta$ value is chosen $(\beta_e =0.01\%)$. From the toroidal mode number scan shown in Fig.~\ref{converg_n_scan}, the most unstable mode is $n=25$ and when $n\geq 50$, the ITG modes are stable. For the multiple-$n$ simulations, it is cheaper to perform a ``partial torus'' instead of ``full torus'' .  GEM simulations are normally performed in a ``partial torus''. For example, a $1/2$th ``partial torus'' $(\Delta n=2)$ means that the simulations include modes $n=0,2,4,\dots$ only (in practical simulations, low-$n$ modes with $0<n<5$ are filtered out, as the field solvers assume a high-$n$ approximation). We chose the maximum toroidal number $n_{max}=45$ for $\Delta  n=15,9,5,3$ and $n_{max}=44$ for $\Delta n=2$. The convergence results are shown in Fig.~\ref{converg_tur_j_rms}. The turbulence time evolution is converged when $\Delta n\le5$ as Fig.~\ref{converg_tur_j_rms}(a) shows. $j_{rms}=\sqrt{\sum_ij_{\parallel e}^2(r_i)\cdot r_i/\sum_i r_i}$ is the root mean square of $j_{\parallel e}$ (the annulus volume at each radial location is approximately $\propto r$) which measures the magnitude of turbulence induced current.  Figure~\ref{converg_tur_j_rms}$(b)$ shows the root mean square of the spontaneous current $j_{rms}$ in ``partial torus'' and ``full torus''. It is converged to a fixed value when $\Delta n\le 5$. The time averaged current profiles in Fig.~\ref{j_multip_mode} show that simulation results of $1/5$th torus are a reasonable approximation of the ``full torus''.

In linear stage, all modes exist and the unstable eigenmodes grow up from the initial noise. Nonlinearly, the most unstable mode is saturated firstly with the generation of zonal 
flows\cite{Diamond_2005}, and the eigenmodes with lower growth rate subseque developed and saturated (Fig.~\ref{mode_spectra}).

The radial profiles of the time averaged spontaneous current over the saturation stage are shown in Fig.~\ref{j_multip_mode}. 
The blue line is the multiple toroidal mode simulation results which includes $n=0,5,10,\dots,45$, the red line includes $n=0,6,9,\dots,45$, the yellow line includes $n=0,6,8,\dots,44$, the purple line includes $0,5,6,\dots,45$ and the black line denotes the reference bootstrap current. The radial profiles of these four cases are not exactly the same, but they all have peaks near the rational mode surfaces $q=1(r/a=0.270)$ and $q=2(r/a=0.705)$. The fine structures can be observed clearly at the exact resonance surface $q=5/5, 6/5(r/a=0.402), 9/5(r/a=0.642), 10/5$  for the  $1/5$th torus simulation,  $q=3/3, 4/3, 5/3, 6/3$ for the $1/3$th torus simulation, $q=2/2, 3/2(r/a=0.537), 4/2$ for the half torus simulation and  $q=1/1, 2/2$ for the ``full torus'' simulation. The ITG harmonic $(n,m)$ is located at $q(r)=m/n$ and the parallel wave number is $k_{\parallel}=(m-nq(r))/qR_0$ which has opposite direction at two sides of $q(r)=m/n$. The turbulence induced current is the total contributions of each harmonic. At the exact resonance surface, the resonant harmonic has much bigger net contribution to the turbulence induced current and the contribution from all harmonics leads to the net current profiles. For example, at the $q=4/3, 5/3$ surfaces, the fine structures can be observed clearly only for the $1/3$th torus cases, but are less significant for the $1/5$th and $1/2$th torus cases since $q=4/3, 5/3$ are not their exact resonance surfaces. As more toroidal harmonics are included, the contribution of harmonics on turbulence induced current is canceled partially and the current fine structures become less visible for most rational mode surfaces.\cite{waltz2006,dominski2017} The turbulence induced current at $q=1,2$ is not significantly reduced when more harmonics are included, because $q=1,2$ surfaces are rational surfaces for each toroidal harmonic. The peaks corrugate the bootstrap current profile, especially in the outer region $r/a>0.5$. For this cyclone base case with $\beta_e=0.01\%$, the magnitude of the turbulence driven current $(\Delta n=1)$ is about $1.4$ times of the bootstrap current, namely, $j_{\parallel e}/j_{bs}\sim 1.4$ at $r/a=0.705$ $(q=2)$.

In the low $\beta$ limit, two cases with $\beta_1=0.01\%$ and $\beta_2=0.03\%$ are performed using $1/5$th torus (by varying density). The ITG growth rate\cite{gorler2016} and the the magnitude of the turbulence intensity (as shown in Fig.~\ref{Beta_scan_j_paral}(a)) of these two cases are almost the same. The turbulence induced current is simply proportional to the density as shown in Fig.~\ref{Beta_scan_j_paral}(b) where the amplitude of normalized current $j_{\parallel e}/en_{ref}c_s$ are similar. According to the amplitude of the turbulence induced current and bootstrap current in the outer region $(r/a>0.5)$ of Fig.~\ref{j_multip_mode}, in the electrostatic limit, the turbulence induced current also corrugates the bootstrap current density profile significantly  at the resonance surface in higher plasma density devices.

\section{Conclusions}
\label{conclusions}
In this work, the current driven by the ion temperature gradient turbulence has been studied using the gyrokinetic code GEM in the collisionless limit.
The principal results of this work are summarized as follows:

\begin{itemize}
    \item[(1)] 
    On account of the high charge-to-mass ratio of electrons, turbulence corrugates the current density profile through changing the electron parallel   
    momentum transport by two mechanisms. One is the electron momentum flux $\Pi_{\parallel e}$ and the other is the electron-ion momentum exchange
    $M_{\parallel e}$. 
    \item[(2)]
    	In linear stage, $\Pi_{\parallel e}$ is proportional to the gradient of turbulence intensity. It traverses the rational surface on  few $\rho_i$ scale and $\nabla \cdot \Pi_{\parallel e}$ has peaks at these surfaces.
    \item[(3)]
     Radial profile and correlation analyses show that the divergence of $\Pi_{\parallel e}$ has the dominant contribution to the current 
    generation over $M_{\parallel e}$ with the fine structure of $\Pi_{\parallel e}$ in the vicinity of rational surface.
     \item[(4)]
     The multiple mode simulation results show that the ITG turbulence driven current has peaks on the rational surface which corrugate the reference bootstrap current especially in the outer region. In electrostatic limit, the magnitude of turbulence induced current increase in higher plasma density tokamaks. 
\end{itemize}

In this work, we focused on the electrostatic ITG modes, and the current driven by trapped electron modes (TEMs) also merits more effort in our future work. Previous work on ion intrinsic rotation demonstrates the rotation reversal as the turbulence changes from ITG modes to TEMs\cite{lu2015,Camenen2011,rice2011}. The current drive efficiency of TEMs can also be different considering its different mode structure and frequency properties compared with ITG modes. Beside the TEMs in the electrostatic limit, the electromagnetic effects or other electromagnetic modes such as KBMs, can also bring new features of the current drive and will be studied in our future works. 

\begin{acknowledgments}
This work is supported by the National Natural Science Foundation of China (Grant Nos. 11822505, 11835016, and
11675257), the Youth Innovation Promotion Association CAS, the Users with Excellence Program of Hefei Science Center CAS (Grant No. 2019HSC-UE013), the Fundamental Research Funds for the Central Universities (Grant No. WK3420000008), and the Collaborative Innovation Program of Hefei Science Center CAS (Grant No. 2019HSC-CIP014).
The numerical calculations in this paper were performed on 
the ShenMa High Performance Computing Cluster in Institute of Plasma Physics, Chinese Academy of Sciences and Hefei advanced computing center.
\end{acknowledgments}

\section*{data availability}

The data that support the findings of this study are available from the corresponding author upon reasonable request.

\clearpage

\setcounter{figure}{0}
\clearpage

\begin{figure}[htbp]
\centering
\includegraphics[scale=0.18]{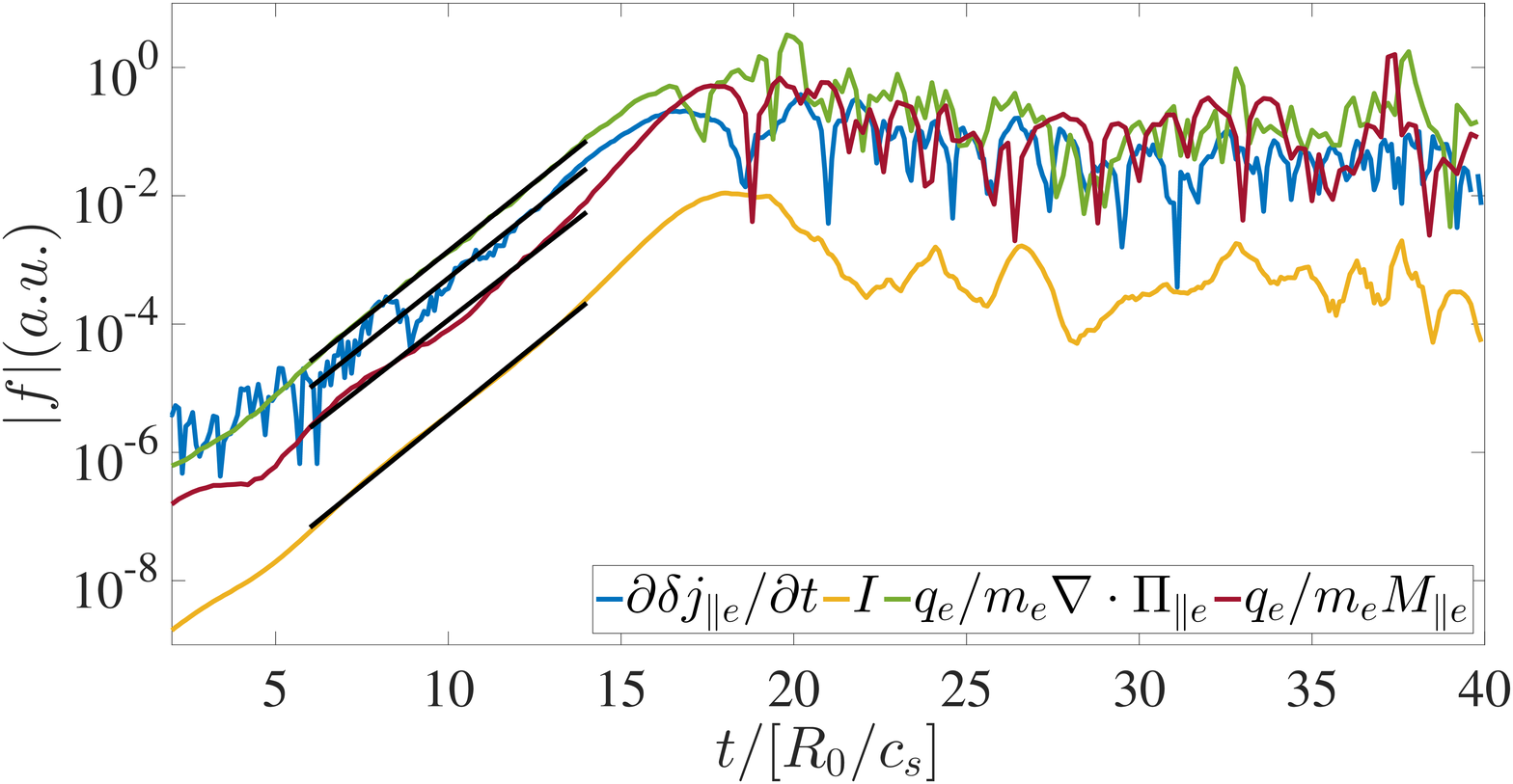}
\caption{The absolute value of physical quantities changes over time, the variation of current $\partial j_{\parallel e}/\partial t$ (blue line), turbulence intensity $I$ (yellow line), the divergence of electron momentum flux $\nabla\cdot\Pi_{\parallel e}$ (green line) and electron-ion momentum exchange $M_{\parallel e}$ (red line) at fixed minor radius where $q(r)=1.3$, for single mode ($n=0, 20$) and low $\beta$ ($\beta=0.1\%$) simulation. After saturation, $\partial j_{\parallel e}/\partial t$ at a radial location changes its sign during its evolution.  \label{Physical_quantities_with_t}}
\end{figure}

\begin{figure}[htbp]
\centering
\subfigure{
\centering
\begin{overpic}
[scale=0.15]{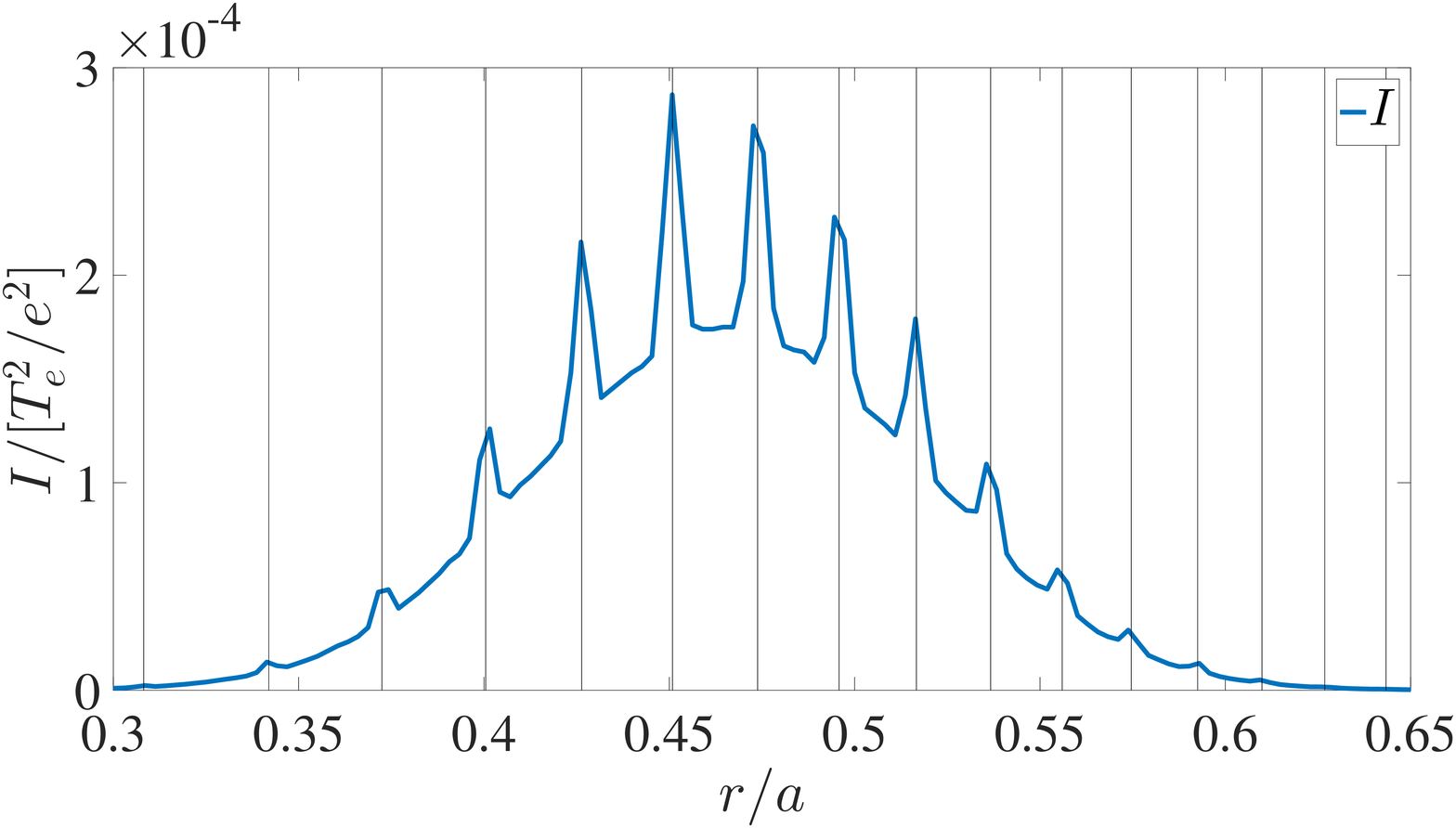}
\put(83,10){$\left(a\right)$}
\end{overpic}
}
\subfigure{
\centering
\begin{overpic}
[scale=0.15]{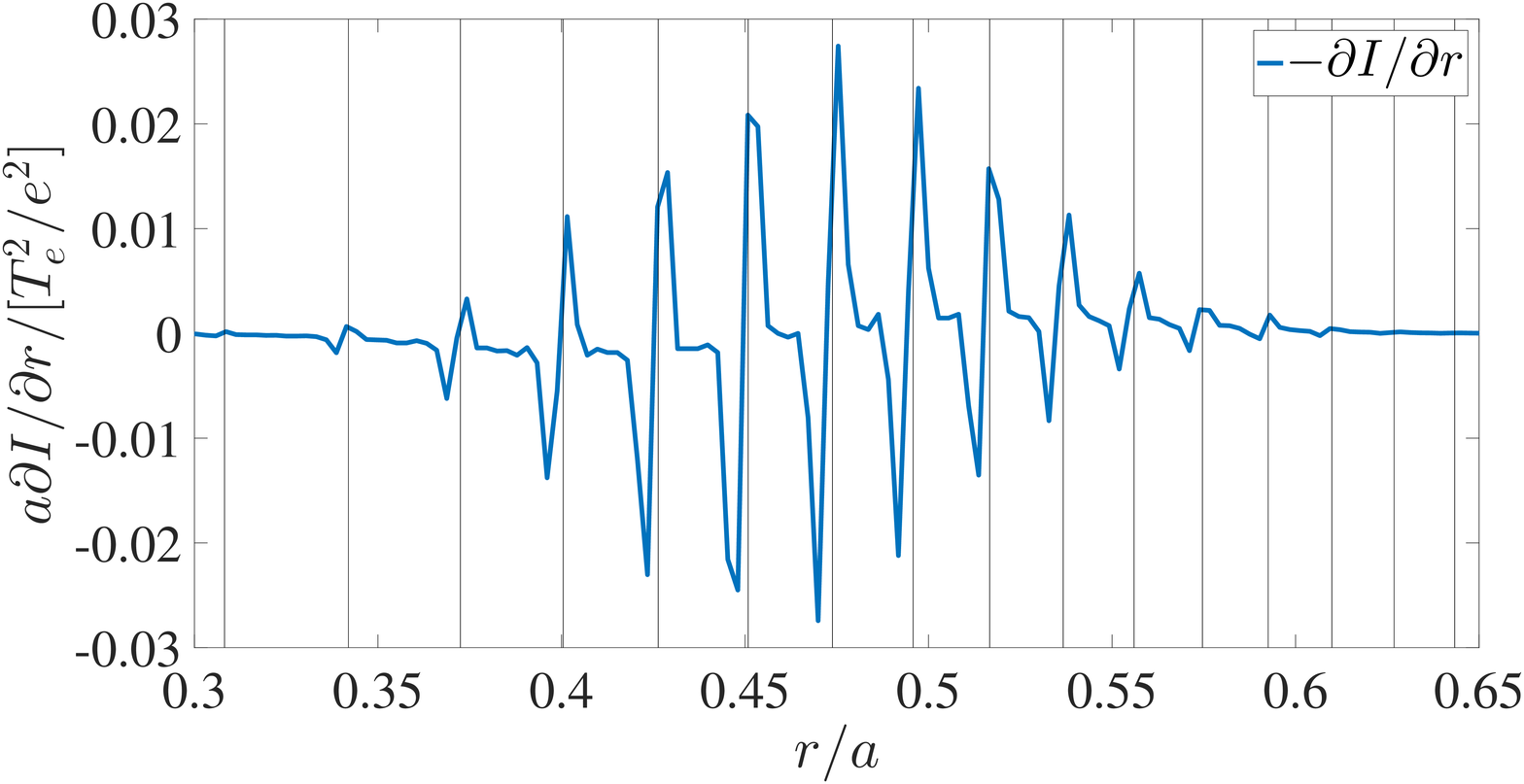}
\put(83,10){$\left(b\right)$}
\end{overpic}
}
\subfigure{
\centering
\begin{overpic}
[scale=0.15]{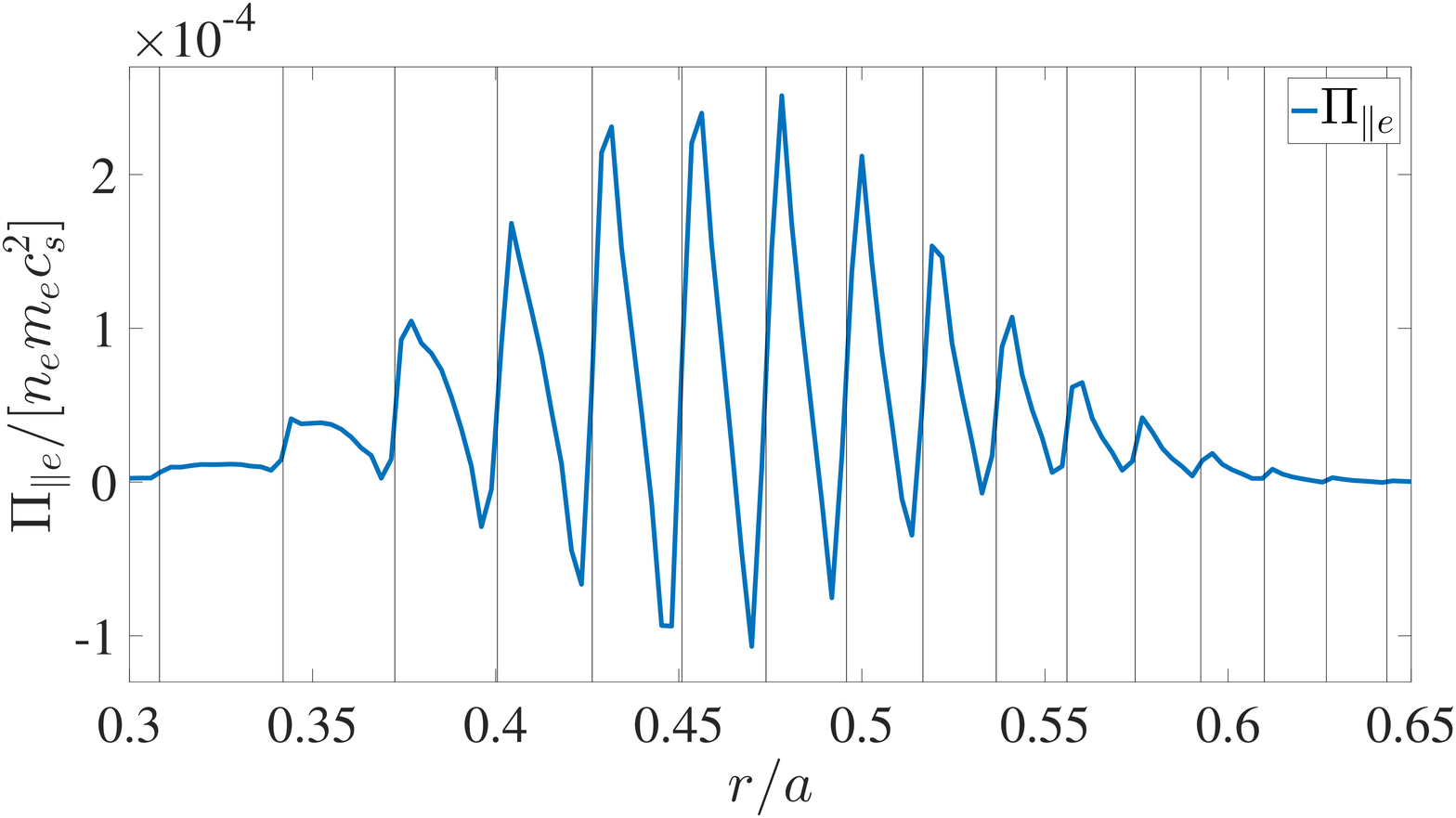}
\put(83,10){$\left(c\right)$}
\end{overpic}
}
\subfigure{
\centering
\begin{overpic}
[scale=0.15]{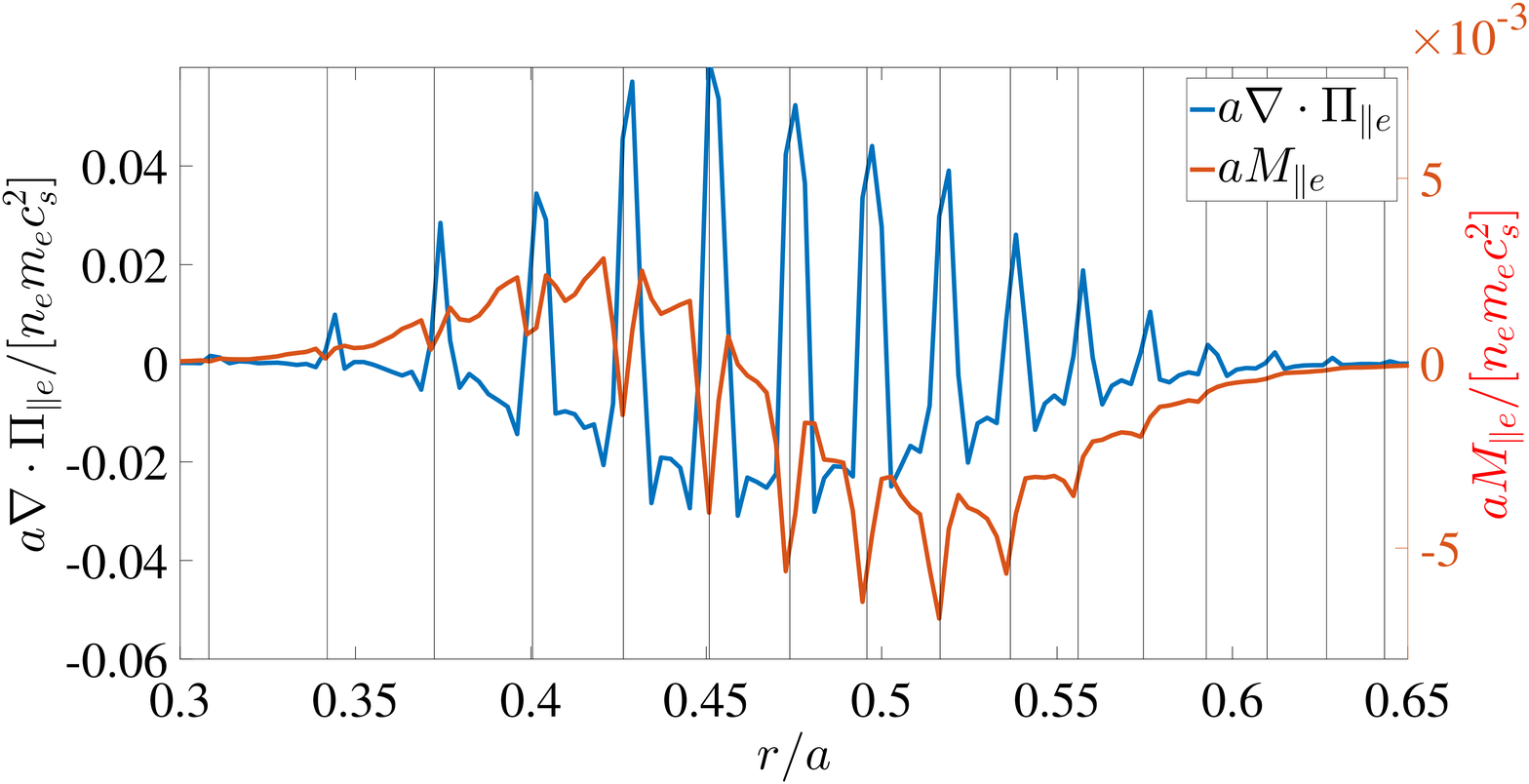}
\put(83,10){$\left(d\right)$}
\end{overpic}
}
\caption{Radial profile of fluctuation intensity $I$ $\left(a\right)$, the radial gradient of fluctuation intensity $\partial I/\partial r$ $\left(b\right)$, the electron momentum flux $\Pi_{\parallel e}$ $\left(c\right)$, the divergence of electron momentum flux $\nabla\cdot\Pi_{\parallel e}$ (blue line) (d) and the turbulence induced electron ion momentum exchange $M_{\parallel e}$ (red line) $\left(d\right)$ at $t=14R_0/c_s$. The vertical black lines represent rational surfaces at $r_s=m/n$ and $m, n$ are poloidal and toroidal mode numbers.}
\label{figure9}
\end{figure}

\begin{figure}[htbp]
\centering
\includegraphics[scale=0.18]{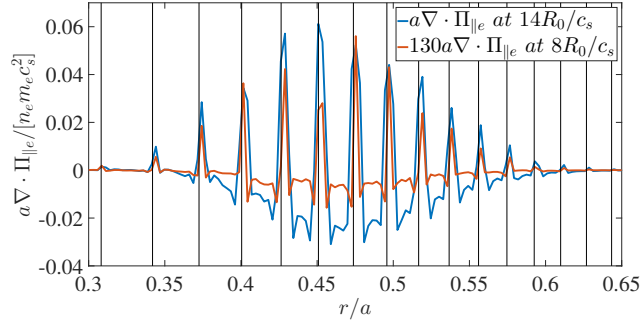}
\caption{Radial profile of the divergence of electron parallel momentum flux $\nabla \cdot\Pi_{\parallel e}$ at different times.} 
\label{figure2}
\end{figure}

\begin{figure}[htbp]
\centering
\subfigure{
\centering
\begin{overpic}
[scale=0.15]{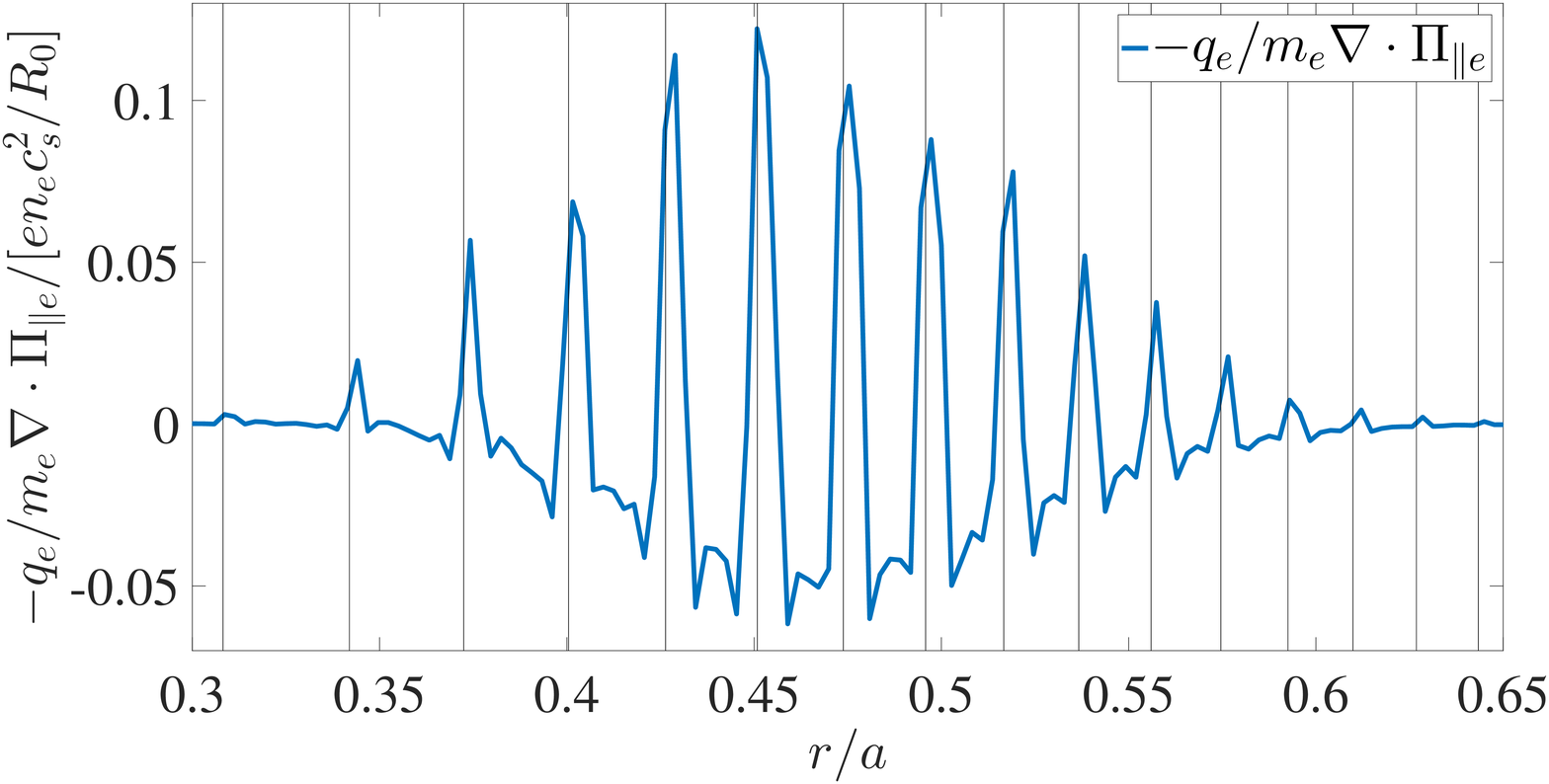}
\put(83,10){$\left(a\right)$}
\end{overpic}
}
\subfigure{
\centering
\begin{overpic}
[scale=0.15]{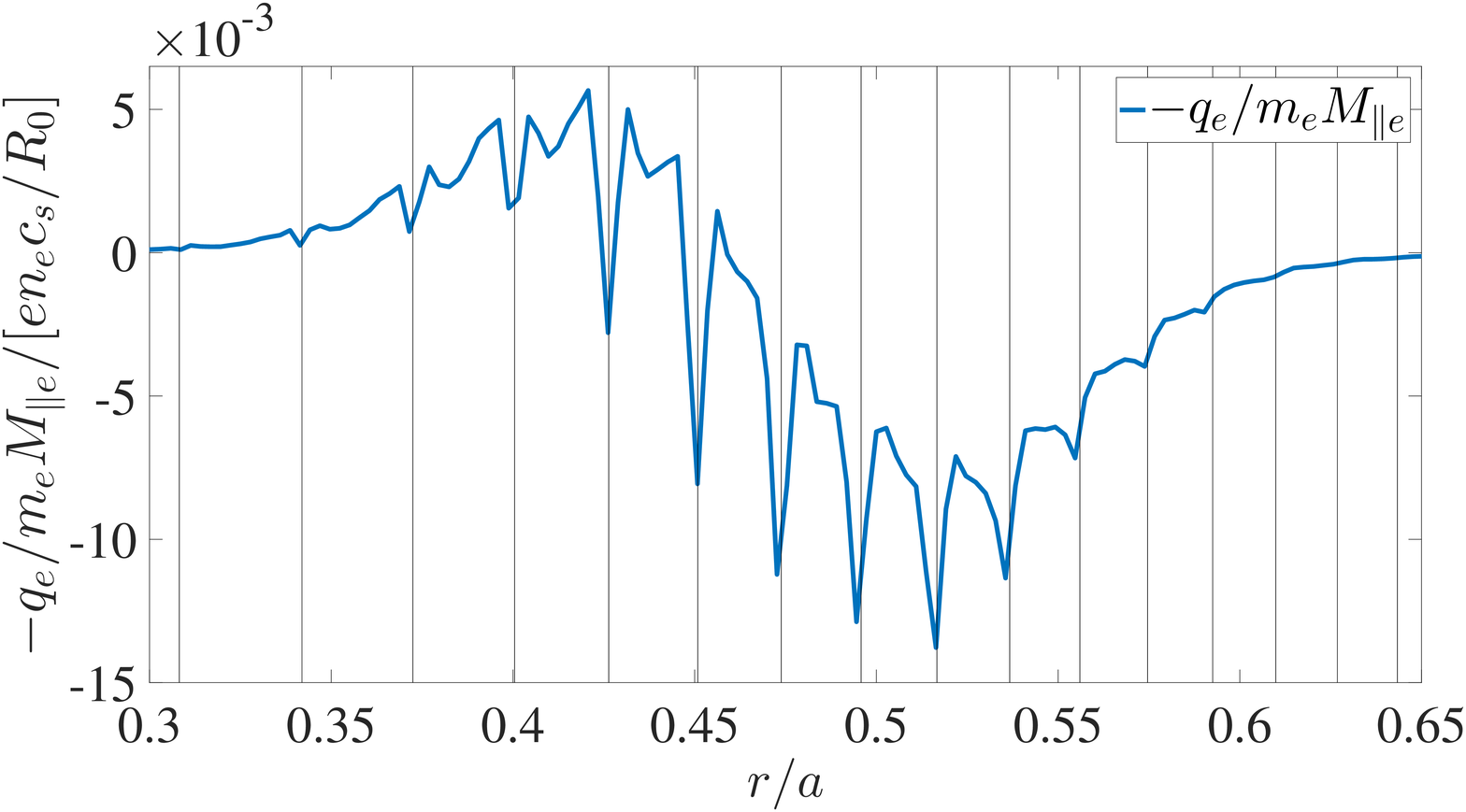}
\put(83,10){$\left(b\right)$}
\end{overpic}
}
\subfigure{
\centering
\begin{overpic}
[scale=0.15]{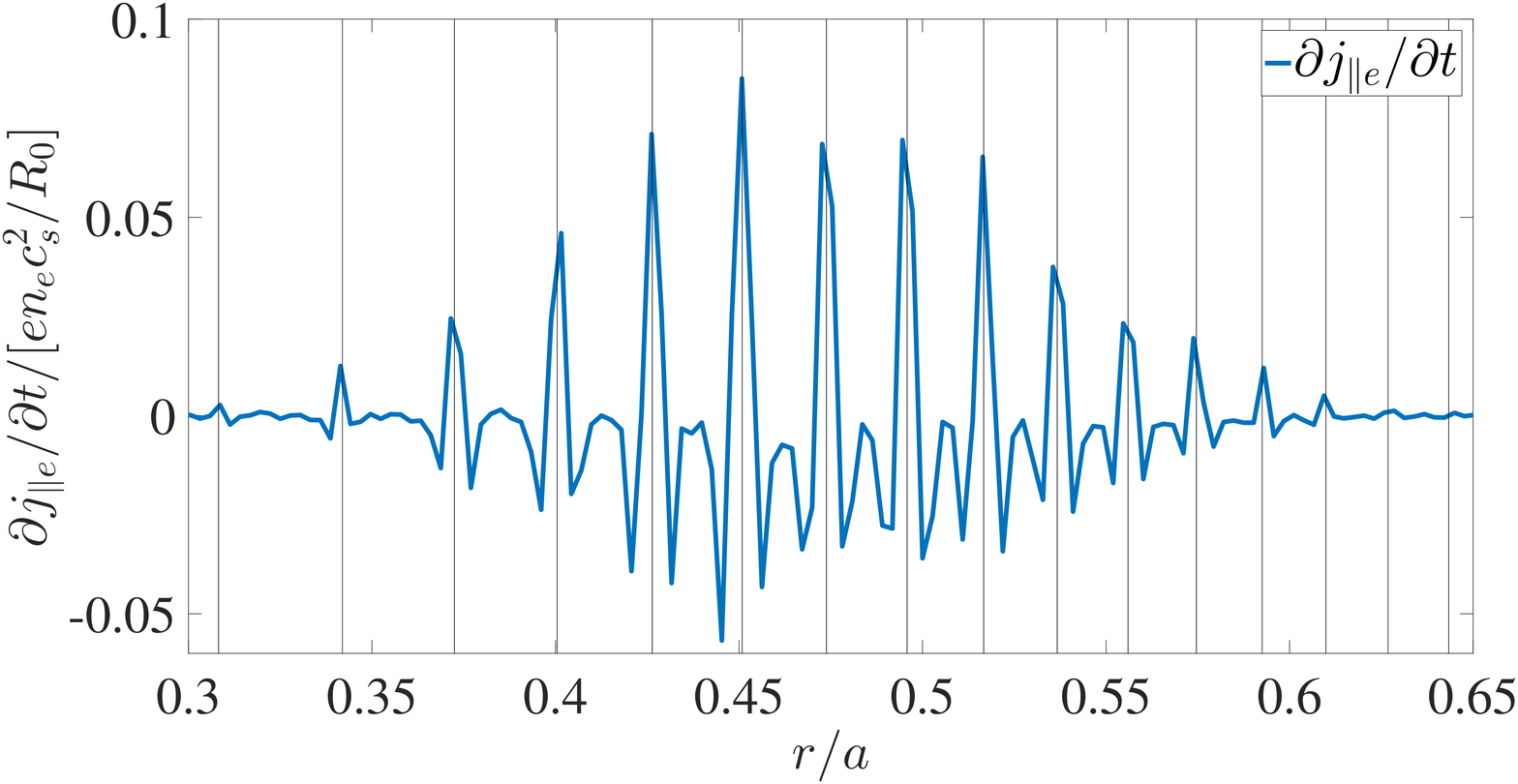}
\put(83,10){$\left(c\right)$}
\end{overpic}
}
\subfigure{
\centering
\begin{overpic}
[scale=0.15]{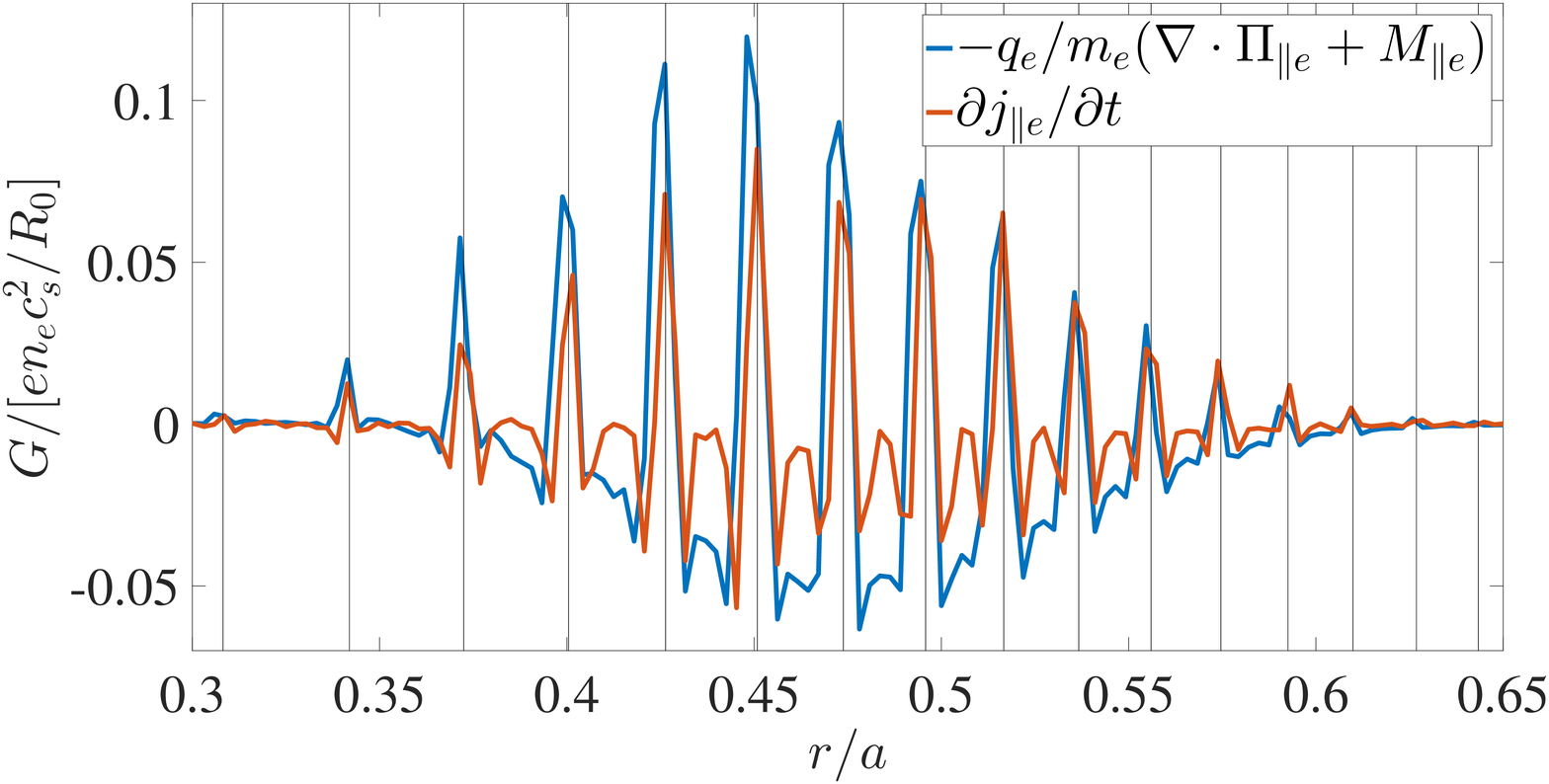}
\put(83,10){$\left(d\right)$}
\end{overpic}
}

\caption{Flux surface averaged results at $t=14R_0/c_s$.
$\left(a\right)$ is the divergence of electron momentum flux ${\nabla\cdot\Pi_{\parallel e}}$,
$\left(b\right)$ is the electron-ion momentum exchange ${M_{\parallel}}$,
$\left(c\right)$ is the turbulence driven current variation with time $\partial j_{\parallel e}/\partial t$  and
$\left(d\right)$ denotes the comparison between the total contribution from the divergence of electron momentum flux and electron-ion momentum exchange (blue line)  and the current variation of time (red line).}
\label{figure3}
\end{figure}

\begin{figure}[t]
\centering
\includegraphics[scale=0.5]{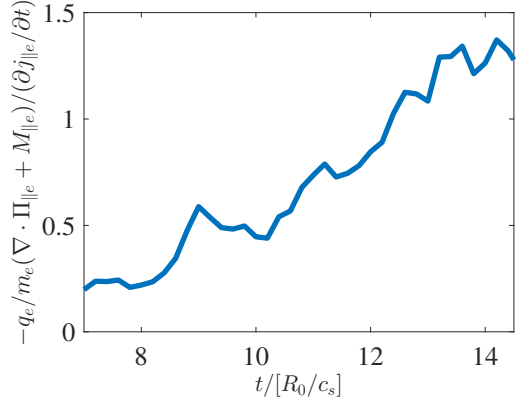}
\caption{Temporal evolution of the ratio between $-q_e/m_e\left(\nabla\cdot\Pi_{\parallel e}+M_{\parallel e}\right)$ and $\left(\partial j/\partial t\right)$ at the rational surface with $q=1.3$.}
\label{Ratio_pJpt_M_Pi}
\end{figure}

\begin{figure}[t]
\centering
\includegraphics[scale=0.18]{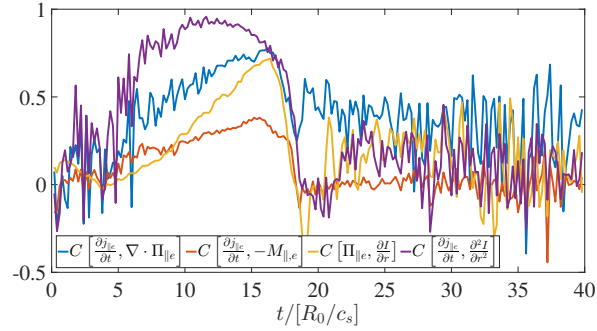}
\caption{The correlation among the variation of turbulence induced current $\partial j_{\parallel e}/\partial t$, the divergence of electron momentum flux $\nabla\cdot\Pi_{\parallel e}$, electron-ion momentum exchange $M_{\parallel e}$, the turbulence intensity gradient $\partial I/\partial r$  and the 
radial second-order derivative of turbulence intensity $\partial^2 I/\partial r^2$.} 
\label{correlation_pJpt_M_Pi}

\end{figure}

\begin{figure}[t]
\centering
\includegraphics[scale=0.18]{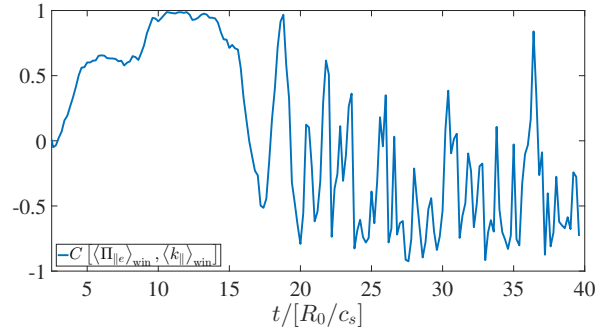}
\caption{The correlation of the windows average results between the electron momentum flux $\Pi_{\parallel e}$ and the spectrum weighted parallel wave vector $\left\langle k_{\parallel}\delta\phi^2\right\rangle$.} 
\label{correlation_Pi_k_paral}

\end{figure}

\begin{figure}[t]
\centering
\includegraphics[scale=0.28]{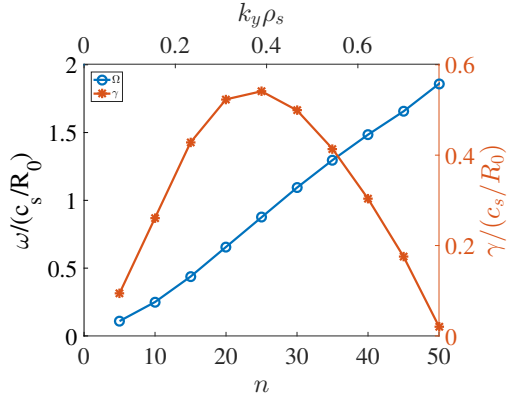}
\caption{The toroidal mode number scan of CBC in lower electron temperature gradient $(\kappa_{Te}=2.23)$ and $\beta$ $(\beta_e=0.01\%)$ limit, the blue line is the frequency and the red line denotes the growth rate. }
\label{converg_n_scan}
\end{figure}

\begin{figure}[htbp]
\centering
\subfigure
{
\begin{minipage}[b]{.45\linewidth}
\centering
\includegraphics[scale=0.175]{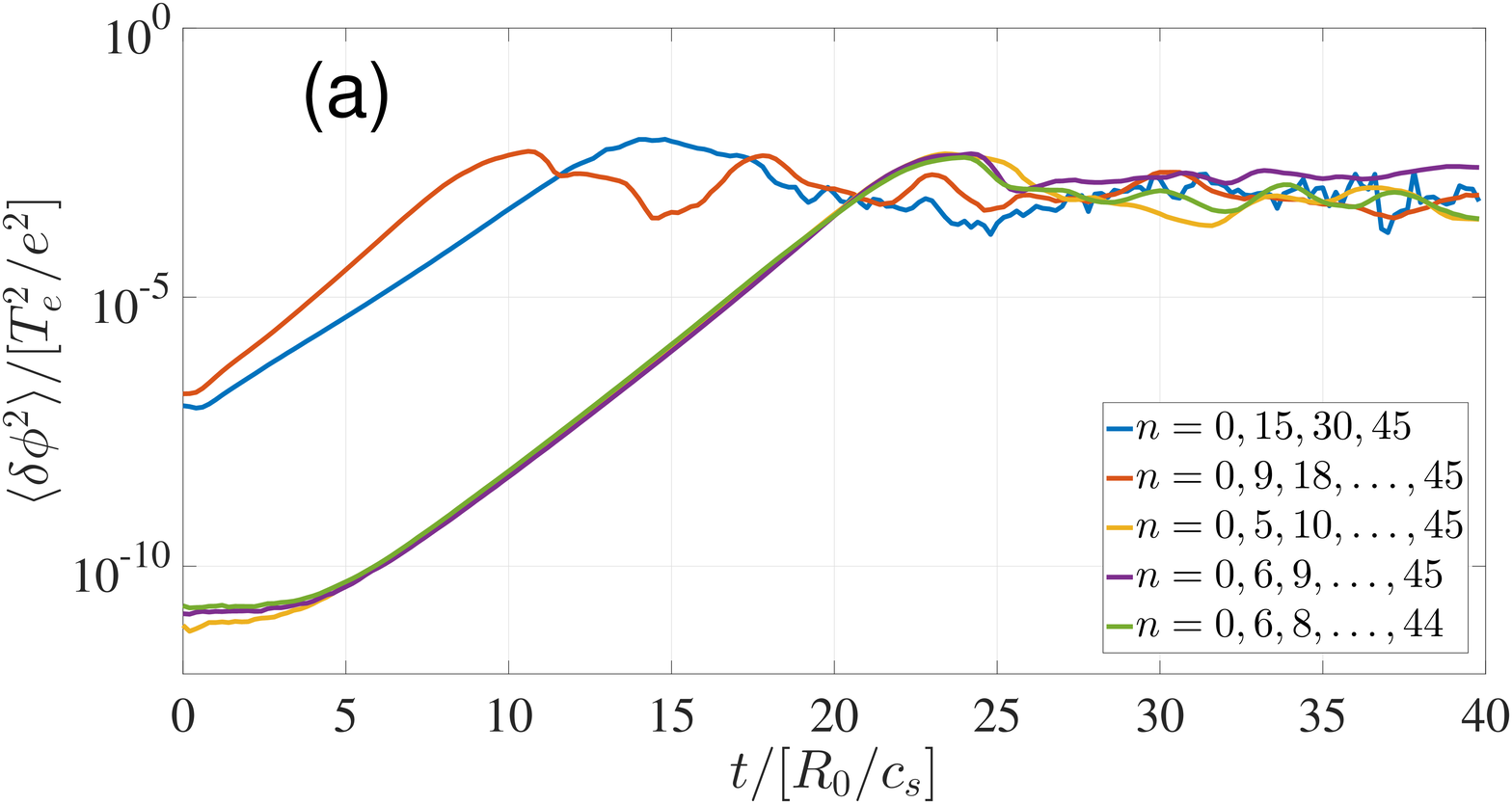}
\end{minipage}
}
\subfigure
{
\begin{minipage}[b]{.45\linewidth}
\centering
\includegraphics[scale=0.4]{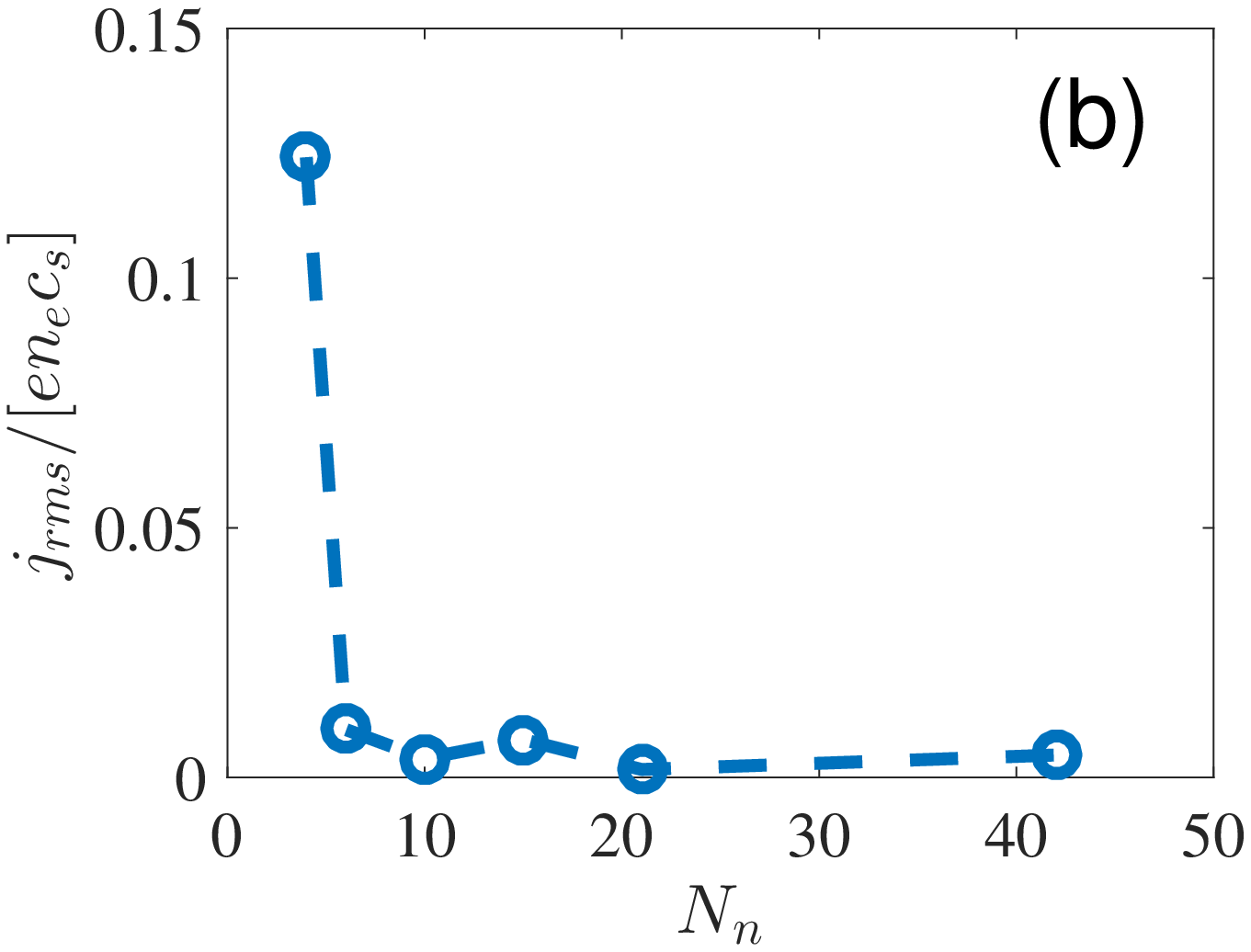}
\end{minipage}
}
\caption{The convergence of the turbulence time evolution $(a)$ and $j_{rms}$ $(b)$ (root mean square of $j_{\parallel e}$) with different ``partial torus'' $\Delta n=15,9,5,3,2$ namely, $N_n=4,6,10,15,21$, $N_n$ is the total toroidal mode number included in the simulations. The $j_{rms}$ of $\Delta n=1, N_n=42$ is also included in (b). 
\label{converg_tur_j_rms}}

\end{figure}

\begin{figure}[t]
\centering
\includegraphics[scale=0.3]{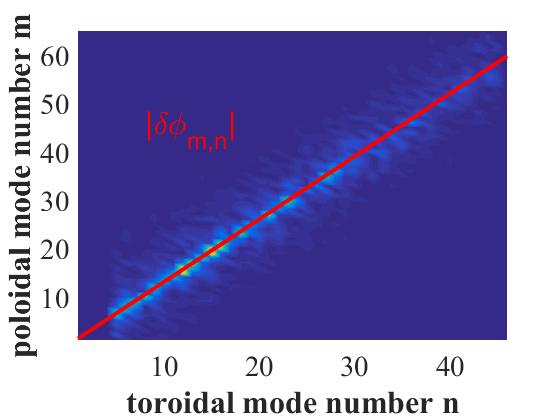}
\caption{Mode spectra $\lvert \delta\phi_{m,n}(r)\rvert$ of ``full torus'' simulation $(\Delta n=1, n=0,5,6,\dots,45)$  at $r/a=0.45$ ($q=1.3$)  in nonlinear stage , the slope of the red line is equal to $m/n=1.3$.}
\label{mode_spectra}

\end{figure}

\begin{figure}[t]
\centering
\includegraphics[scale=0.18]{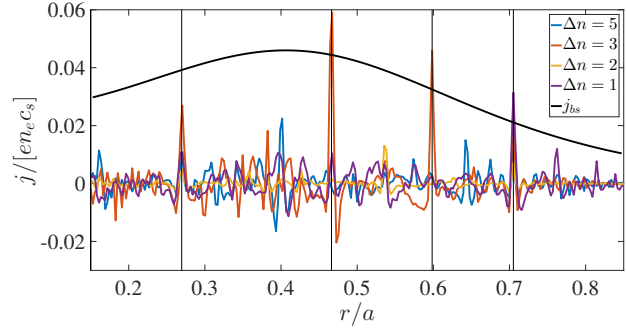}
\caption{The radial profile of the time averaged turbulence driven current during the turbulence saturation stage with multiple mode simulations $(\Delta n=5, n=0,5,10,\dots,45)$ (blue line), $(\Delta n=3, n=0,6,9,\dots,45)$ (red line), $(\Delta n=2, n=0,6,8,\dots, 44)$ (yellow line), $(\Delta n=1, n=0,5,6,\dots,45)$ (purple line) and the solid black line denotes the reference neoclassical bootstrap current $j_{bs}$. The vertical lines are the rational surfaces of $n=3$  $(q=3/3, 4/3, 5/3, 6/3, 7/3)$.}
\label{j_multip_mode}

\end{figure}

\begin{figure}[htbp]
\centering
\subfigure
{
\begin{minipage}[b]{.55\linewidth}
\centering
\includegraphics[scale=0.18]{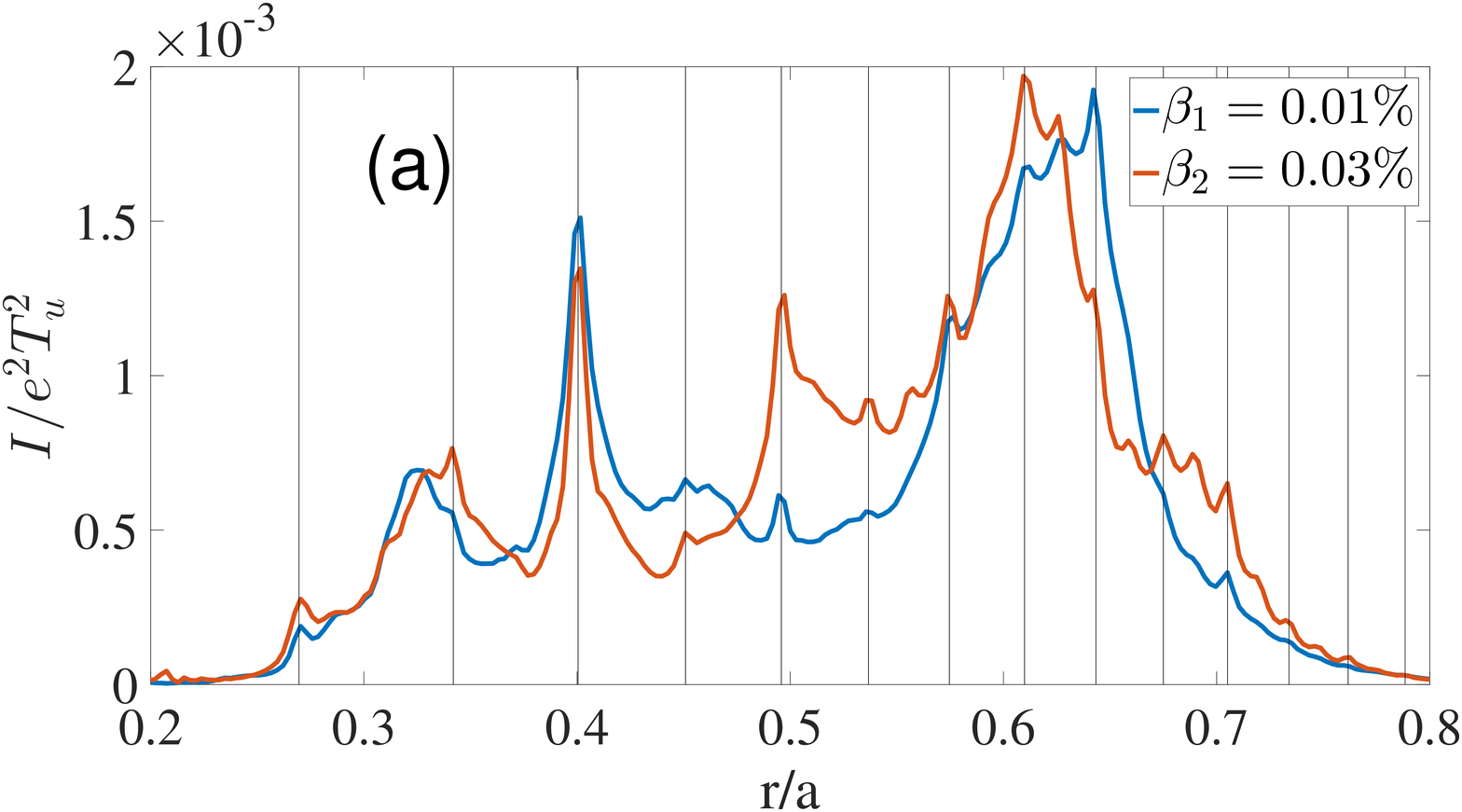}
\end{minipage}
}
\subfigure
{
\begin{minipage}[b]{.55\linewidth}
\centering
\includegraphics[scale=0.18]{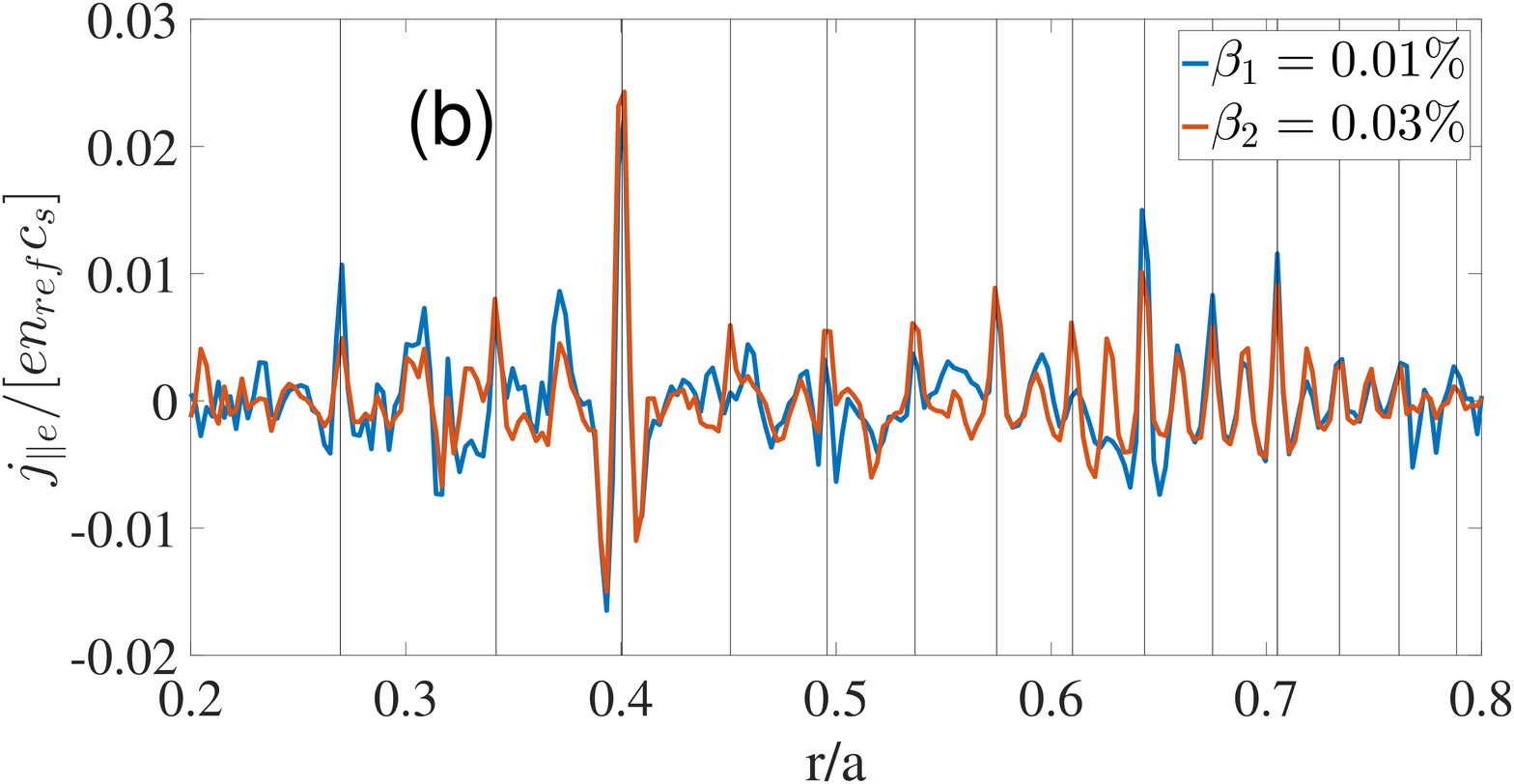}
\end{minipage}
}
\caption{Time averaged turbulence intensity $(a)$ and turbulence induced current $(b)$ with different $\beta_e$ value (by varying density, $\beta_1/\beta_2=n_{ref1}/n_{ref2}=1/3$).}
\label{Beta_scan_j_paral}
\end{figure}


\begin{thebibliography}{35}%
\makeatletter
\providecommand \@ifxundefined [1]{%
 \@ifx{#1\undefined}
}%
\providecommand \@ifnum [1]{%
 \ifnum #1\expandafter \@firstoftwo
 \else \expandafter \@secondoftwo
 \fi
}%
\providecommand \@ifx [1]{%
 \ifx #1\expandafter \@firstoftwo
 \else \expandafter \@secondoftwo
 \fi
}%
\providecommand \natexlab [1]{#1}%
\providecommand \enquote  [1]{``#1''}%
\providecommand \bibnamefont  [1]{#1}%
\providecommand \bibfnamefont [1]{#1}%
\providecommand \citenamefont [1]{#1}%
\providecommand \href@noop [0]{\@secondoftwo}%
\providecommand \href [0]{\begingroup \@sanitize@url \@href}%
\providecommand \@href[1]{\@@startlink{#1}\@@href}%
\providecommand \@@href[1]{\endgroup#1\@@endlink}%
\providecommand \@sanitize@url [0]{\catcode `\\12\catcode `\$12\catcode
  `\&12\catcode `\#12\catcode `\^12\catcode `\_12\catcode `\%12\relax}%
\providecommand \@@startlink[1]{}%
\providecommand \@@endlink[0]{}%
\providecommand \url  [0]{\begingroup\@sanitize@url \@url }%
\providecommand \@url [1]{\endgroup\@href {#1}{\urlprefix }}%
\providecommand \urlprefix  [0]{URL }%
\providecommand \Eprint [0]{\href }%
\providecommand \doibase [0]{http://dx.doi.org/}%
\providecommand \selectlanguage [0]{\@gobble}%
\providecommand \bibinfo  [0]{\@secondoftwo}%
\providecommand \bibfield  [0]{\@secondoftwo}%
\providecommand \translation [1]{[#1]}%
\providecommand \BibitemOpen [0]{}%
\providecommand \bibitemStop [0]{}%
\providecommand \bibitemNoStop [0]{.\EOS\space}%
\providecommand \EOS [0]{\spacefactor3000\relax}%
\providecommand \BibitemShut  [1]{\csname bibitem#1\endcsname}%
\let\auto@bib@innerbib\@empty
\bibitem [{\citenamefont {Bickrton}, \citenamefont {Connor},\ and\
  \citenamefont {Taylor}(1971)}]{bickerton1971}%
  \BibitemOpen
  \bibfield  {author} {\bibinfo {author} {\bibfnamefont {R.~J.}\ \bibnamefont
  {Bickrton}}, \bibinfo {author} {\bibfnamefont {J.~W.}\ \bibnamefont
  {Connor}}, \ and\ \bibinfo {author} {\bibfnamefont {J.~B.}\ \bibnamefont
  {Taylor}},\ }\href@noop {} {\bibfield  {journal} {\bibinfo  {journal} {Nat.
  Phys. Sci.}\ }\textbf {\bibinfo {volume} {229}},\ \bibinfo {pages} {110}
  (\bibinfo {year} {1971})}\BibitemShut {NoStop}%
\bibitem [{\citenamefont {Peeters}(2000)}]{peeters2000}%
  \BibitemOpen
  \bibfield  {author} {\bibinfo {author} {\bibfnamefont {A.~G.}\ \bibnamefont
  {Peeters}},\ }\href@noop {} {\bibfield  {journal} {\bibinfo  {journal}
  {Plasma Phys. Controlled Fusion}\ }\textbf {\bibinfo {volume} {42}},\
  \bibinfo {pages} {B231} (\bibinfo {year} {2000})}\BibitemShut {NoStop}%
\bibitem [{\citenamefont {McDevitt}, \citenamefont {Tang},\ and\ \citenamefont
  {Guo}(2017)}]{mcdevitt2017}%
  \BibitemOpen
  \bibfield  {author} {\bibinfo {author} {\bibfnamefont {C.~J.}\ \bibnamefont
  {McDevitt}}, \bibinfo {author} {\bibfnamefont {X.~Z.}\ \bibnamefont {Tang}},
  \ and\ \bibinfo {author} {\bibfnamefont {Z.~H.}\ \bibnamefont {Guo}},\
  }\href@noop {} {\bibfield  {journal} {\bibinfo  {journal} {Phys. Plasmas}\
  }\textbf {\bibinfo {volume} {24}},\ \bibinfo {pages} {082307} (\bibinfo
  {year} {2017})}\BibitemShut {NoStop}%
\bibitem [{\citenamefont {Wang}\ \emph {et~al.}(2019)\citenamefont {Wang},
  \citenamefont {Hahm}, \citenamefont {Startsev}, \citenamefont {Ethier},
  \citenamefont {Chen}, \citenamefont {Yoo},\ and\ \citenamefont
  {Ma}}]{wang2019}%
  \BibitemOpen
  \bibfield  {author} {\bibinfo {author} {\bibfnamefont {W.~X.}\ \bibnamefont
  {Wang}}, \bibinfo {author} {\bibfnamefont {T.~S.}\ \bibnamefont {Hahm}},
  \bibinfo {author} {\bibfnamefont {E.~A.}\ \bibnamefont {Startsev}}, \bibinfo
  {author} {\bibfnamefont {S.}~\bibnamefont {Ethier}}, \bibinfo {author}
  {\bibfnamefont {J.}~\bibnamefont {Chen}}, \bibinfo {author} {\bibfnamefont
  {M.~G.}\ \bibnamefont {Yoo}}, \ and\ \bibinfo {author} {\bibfnamefont
  {C.~H.}\ \bibnamefont {Ma}},\ }\href@noop {} {\bibfield  {journal} {\bibinfo
  {journal} {Nucl. Fusion}\ }\textbf {\bibinfo {volume} {59}},\ \bibinfo
  {pages} {084002} (\bibinfo {year} {2019})}\BibitemShut {NoStop}%
\bibitem [{\citenamefont {Cai}(2018)}]{cai2018}%
  \BibitemOpen
  \bibfield  {author} {\bibinfo {author} {\bibfnamefont {H.~S.}\ \bibnamefont
  {Cai}},\ }\href@noop {} {\bibfield  {journal} {\bibinfo  {journal} {Nucl.
  Fusion}\ }\textbf {\bibinfo {volume} {59}},\ \bibinfo {pages} {026009}
  (\bibinfo {year} {2018})}\BibitemShut {NoStop}%
\bibitem [{\citenamefont {Itoh}\ and\ \citenamefont {Itoh}(1988)}]{ITOH1988}%
  \BibitemOpen
  \bibfield  {author} {\bibinfo {author} {\bibfnamefont {S.-I.}\ \bibnamefont
  {Itoh}}\ and\ \bibinfo {author} {\bibfnamefont {K.}~\bibnamefont {Itoh}},\
  }\href@noop {} {\bibfield  {journal} {\bibinfo  {journal} {Phys. Lett. A}\
  }\textbf {\bibinfo {volume} {127}},\ \bibinfo {pages} {267} (\bibinfo {year}
  {1988})}\BibitemShut {NoStop}%
\bibitem [{\citenamefont {Hinton}, \citenamefont {Waltz},\ and\ \citenamefont
  {Candy}(2004)}]{hinton2004}%
  \BibitemOpen
  \bibfield  {author} {\bibinfo {author} {\bibfnamefont {F.~L.}\ \bibnamefont
  {Hinton}}, \bibinfo {author} {\bibfnamefont {R.~E.}\ \bibnamefont {Waltz}}, \
  and\ \bibinfo {author} {\bibfnamefont {J.}~\bibnamefont {Candy}},\
  }\href@noop {} {\bibfield  {journal} {\bibinfo  {journal} {Phys. Plasmas}\
  }\textbf {\bibinfo {volume} {11}},\ \bibinfo {pages} {2433} (\bibinfo {year}
  {2004})}\BibitemShut {NoStop}%
\bibitem [{\citenamefont {Garbet}\ \emph {et~al.}(2014)\citenamefont {Garbet},
  \citenamefont {Esteve}, \citenamefont {Sarazin}, \citenamefont
  {Dif-Pradalier}, \citenamefont {Ghendrih}, \citenamefont {Grandgirard},
  \citenamefont {Latu},\ and\ \citenamefont {Smolyakov}}]{garbet2014}%
  \BibitemOpen
  \bibfield  {author} {\bibinfo {author} {\bibfnamefont {X.}~\bibnamefont
  {Garbet}}, \bibinfo {author} {\bibfnamefont {D.}~\bibnamefont {Esteve}},
  \bibinfo {author} {\bibfnamefont {Y.}~\bibnamefont {Sarazin}}, \bibinfo
  {author} {\bibfnamefont {G.}~\bibnamefont {Dif-Pradalier}}, \bibinfo {author}
  {\bibfnamefont {P.}~\bibnamefont {Ghendrih}}, \bibinfo {author}
  {\bibfnamefont {V.}~\bibnamefont {Grandgirard}}, \bibinfo {author}
  {\bibfnamefont {G.}~\bibnamefont {Latu}}, \ and\ \bibinfo {author}
  {\bibfnamefont {A.}~\bibnamefont {Smolyakov}},\ }\href@noop {} {\bibfield
  {journal} {\bibinfo  {journal} {J. Phys.: Conf. Ser.}\ }\textbf {\bibinfo
  {volume} {561}},\ \bibinfo {pages} {012007} (\bibinfo {year}
  {2014})}\BibitemShut {NoStop}%
\bibitem [{\citenamefont {Yi}, \citenamefont {Jhang},\ and\ \citenamefont
  {Kwon}(2016)}]{yi2016}%
  \BibitemOpen
  \bibfield  {author} {\bibinfo {author} {\bibfnamefont {S.}~\bibnamefont
  {Yi}}, \bibinfo {author} {\bibfnamefont {H.}~\bibnamefont {Jhang}}, \ and\
  \bibinfo {author} {\bibfnamefont {J.~M.}\ \bibnamefont {Kwon}},\ }\href@noop
  {} {\bibfield  {journal} {\bibinfo  {journal} {Phys. Plasmas}\ }\textbf
  {\bibinfo {volume} {23}},\ \bibinfo {pages} {102514} (\bibinfo {year}
  {2016})}\BibitemShut {NoStop}%
\bibitem [{\citenamefont {He}\ \emph {et~al.}(2018)\citenamefont {He},
  \citenamefont {Wang}, \citenamefont {Peng}, \citenamefont {Guo},\ and\
  \citenamefont {Zhuang}}]{he2018}%
  \BibitemOpen
  \bibfield  {author} {\bibinfo {author} {\bibfnamefont {W.}~\bibnamefont
  {He}}, \bibinfo {author} {\bibfnamefont {L.}~\bibnamefont {Wang}}, \bibinfo
  {author} {\bibfnamefont {S.~T.}\ \bibnamefont {Peng}}, \bibinfo {author}
  {\bibfnamefont {W.~X.}\ \bibnamefont {Guo}}, \ and\ \bibinfo {author}
  {\bibfnamefont {G.}~\bibnamefont {Zhuang}},\ }\href@noop {} {\bibfield
  {journal} {\bibinfo  {journal} {Nucl. Fusion}\ }\textbf {\bibinfo {volume}
  {58}},\ \bibinfo {pages} {106004} (\bibinfo {year} {2018})}\BibitemShut
  {NoStop}%
\bibitem [{\citenamefont {McDevitt}, \citenamefont {Tang},\ and\ \citenamefont
  {Guo}(2013)}]{McDevitt2013}%
  \BibitemOpen
  \bibfield  {author} {\bibinfo {author} {\bibfnamefont {C.~J.}\ \bibnamefont
  {McDevitt}}, \bibinfo {author} {\bibfnamefont {X.~Z.}\ \bibnamefont {Tang}},
  \ and\ \bibinfo {author} {\bibfnamefont {Z.~H.}\ \bibnamefont {Guo}},\ }\href
  {https://link.aps.org/doi/10.1103/PhysRevLett.111.205002} {\bibfield
  {journal} {\bibinfo  {journal} {Phys. Rev. Lett.}\ }\textbf {\bibinfo
  {volume} {111}},\ \bibinfo {pages} {205002} (\bibinfo {year}
  {2013})}\BibitemShut {NoStop}%
\bibitem [{\citenamefont {Wang}\ \emph {et~al.}(2012)\citenamefont {Wang},
  \citenamefont {Ethier}, \citenamefont {Hinton}, \citenamefont {Hahm},
  \citenamefont {Boozer}, \citenamefont {Diamond}, \citenamefont {Tang},\ and\
  \citenamefont {Li}}]{wang2012}%
  \BibitemOpen
  \bibfield  {author} {\bibinfo {author} {\bibfnamefont {W.~W.}\ \bibnamefont
  {Wang}}, \bibinfo {author} {\bibfnamefont {S.}~\bibnamefont {Ethier}},
  \bibinfo {author} {\bibfnamefont {F.~L.}\ \bibnamefont {Hinton}}, \bibinfo
  {author} {\bibfnamefont {T.~S.}\ \bibnamefont {Hahm}}, \bibinfo {author}
  {\bibfnamefont {A.}~\bibnamefont {Boozer}}, \bibinfo {author} {\bibfnamefont
  {P.~H.}\ \bibnamefont {Diamond}}, \bibinfo {author} {\bibfnamefont {W.~M.}\
  \bibnamefont {Tang}}, \ and\ \bibinfo {author} {\bibfnamefont {Z.~Q.}\
  \bibnamefont {Li}},\ }in\ \href@noop {} {\emph {\bibinfo {booktitle} {24th
  Int. Conf. on Fusion Energy (San Diego, 2012),TH/P7-14}}}\ (\bibinfo {year}
  {2012})\BibitemShut {NoStop}%
\bibitem [{\citenamefont {Waltz}\ \emph {et~al.}(2006)\citenamefont {Waltz},
  \citenamefont {Austin}, \citenamefont {Burrell},\ and\ \citenamefont
  {Candy}}]{waltz2006}%
  \BibitemOpen
  \bibfield  {author} {\bibinfo {author} {\bibfnamefont {R.~E.}\ \bibnamefont
  {Waltz}}, \bibinfo {author} {\bibfnamefont {M.~E.}\ \bibnamefont {Austin}},
  \bibinfo {author} {\bibfnamefont {K.~H.}\ \bibnamefont {Burrell}}, \ and\
  \bibinfo {author} {\bibfnamefont {J.}~\bibnamefont {Candy}},\ }\href
  {\doibase 10.1063/1.2195418} {\bibfield  {journal} {\bibinfo  {journal}
  {Phys. of Plasmas}\ }\textbf {\bibinfo {volume} {13}},\ \bibinfo {pages}
  {052301} (\bibinfo {year} {2006})}\BibitemShut {NoStop}%
\bibitem [{\citenamefont {Dominski}\ \emph {et~al.}(2017)\citenamefont
  {Dominski}, \citenamefont {McMillan}, \citenamefont {Brunner}, \citenamefont
  {Merlo}, \citenamefont {Tran},\ and\ \citenamefont {Villard}}]{dominski2017}%
  \BibitemOpen
  \bibfield  {author} {\bibinfo {author} {\bibfnamefont {J.}~\bibnamefont
  {Dominski}}, \bibinfo {author} {\bibfnamefont {B.~F.}\ \bibnamefont
  {McMillan}}, \bibinfo {author} {\bibfnamefont {S.}~\bibnamefont {Brunner}},
  \bibinfo {author} {\bibfnamefont {G.}~\bibnamefont {Merlo}}, \bibinfo
  {author} {\bibfnamefont {T.~M.}\ \bibnamefont {Tran}}, \ and\ \bibinfo
  {author} {\bibfnamefont {L.}~\bibnamefont {Villard}},\ }\href {\doibase
  10.1063/1.4976120} {\bibfield  {journal} {\bibinfo  {journal} {Phys. of
  Plasmas}\ }\textbf {\bibinfo {volume} {24}},\ \bibinfo {pages} {022308}
  (\bibinfo {year} {2017})}\BibitemShut {NoStop}%
\bibitem [{\citenamefont {Chen}\ and\ \citenamefont {Parker}(2007)}]{chen2007}%
  \BibitemOpen
  \bibfield  {author} {\bibinfo {author} {\bibfnamefont {Y.}~\bibnamefont
  {Chen}}\ and\ \bibinfo {author} {\bibfnamefont {S.~E.}\ \bibnamefont
  {Parker}},\ }\href@noop {} {\bibfield  {journal} {\bibinfo  {journal} {J.
  Comput. Phys.}\ }\textbf {\bibinfo {volume} {220}},\ \bibinfo {pages} {839}
  (\bibinfo {year} {2007})}\BibitemShut {NoStop}%
\bibitem [{\citenamefont {Chen}\ \emph {et~al.}(2010)\citenamefont {Chen},
  \citenamefont {Parker}, \citenamefont {Lang},\ and\ \citenamefont
  {Fu}}]{chen2010}%
  \BibitemOpen
  \bibfield  {author} {\bibinfo {author} {\bibfnamefont {Y.}~\bibnamefont
  {Chen}}, \bibinfo {author} {\bibfnamefont {S.~E.}\ \bibnamefont {Parker}},
  \bibinfo {author} {\bibfnamefont {J.~Y.}\ \bibnamefont {Lang}}, \ and\
  \bibinfo {author} {\bibfnamefont {G.~Y.}\ \bibnamefont {Fu}},\ }\href@noop {}
  {\bibfield  {journal} {\bibinfo  {journal} {Phys. Plasmas}\ }\textbf
  {\bibinfo {volume} {17}},\ \bibinfo {pages} {102504} (\bibinfo {year}
  {2010})}\BibitemShut {NoStop}%
\bibitem [{\citenamefont {Chen}\ and\ \citenamefont {Parker}(2003)}]{chen2003}%
  \BibitemOpen
  \bibfield  {author} {\bibinfo {author} {\bibfnamefont {Y.}~\bibnamefont
  {Chen}}\ and\ \bibinfo {author} {\bibfnamefont {S.~E.}\ \bibnamefont
  {Parker}},\ }\href@noop {} {\bibfield  {journal} {\bibinfo  {journal} {J.
  Comput. Phys.}\ }\textbf {\bibinfo {volume} {189}},\ \bibinfo {pages} {463}
  (\bibinfo {year} {2003})}\BibitemShut {NoStop}%
\bibitem [{\citenamefont {Lee}\ \emph {et~al.}(2001)\citenamefont {Lee},
  \citenamefont {Lewandowski}, \citenamefont {Hahm},\ and\ \citenamefont
  {Lin}}]{lee2001}%
  \BibitemOpen
  \bibfield  {author} {\bibinfo {author} {\bibfnamefont {W.~W.}\ \bibnamefont
  {Lee}}, \bibinfo {author} {\bibfnamefont {J.~L.~V.}\ \bibnamefont
  {Lewandowski}}, \bibinfo {author} {\bibfnamefont {T.~S.}\ \bibnamefont
  {Hahm}}, \ and\ \bibinfo {author} {\bibfnamefont {Z.}~\bibnamefont {Lin}},\
  }\href@noop {} {\bibfield  {journal} {\bibinfo  {journal} {Phys. Plasmas}\
  }\textbf {\bibinfo {volume} {8}},\ \bibinfo {pages} {4435} (\bibinfo {year}
  {2001})}\BibitemShut {NoStop}%
\bibitem [{\citenamefont {Dimits}\ \emph {et~al.}(2000)\citenamefont {Dimits},
  \citenamefont {Bateman}, \citenamefont {Beer}, \citenamefont {Cohen},
  \citenamefont {Dorland}, \citenamefont {Hammett}, \citenamefont {Kim},
  \citenamefont {Kinsey}, \citenamefont {Kotschenreuther}, \citenamefont
  {Kritz}, \citenamefont {Lao}, \citenamefont {Mandrekas}, \citenamefont
  {Nevins}, \citenamefont {Parker}, \citenamefont {Redd}, \citenamefont
  {Shumaker}, \citenamefont {Sydora},\ and\ \citenamefont
  {Weiland}}]{dimits2000}%
  \BibitemOpen
  \bibfield  {author} {\bibinfo {author} {\bibfnamefont {A.~M.}\ \bibnamefont
  {Dimits}}, \bibinfo {author} {\bibfnamefont {G.}~\bibnamefont {Bateman}},
  \bibinfo {author} {\bibfnamefont {M.~A.}\ \bibnamefont {Beer}}, \bibinfo
  {author} {\bibfnamefont {B.~I.}\ \bibnamefont {Cohen}}, \bibinfo {author}
  {\bibfnamefont {W.}~\bibnamefont {Dorland}}, \bibinfo {author} {\bibfnamefont
  {G.~W.}\ \bibnamefont {Hammett}}, \bibinfo {author} {\bibfnamefont
  {C.}~\bibnamefont {Kim}}, \bibinfo {author} {\bibfnamefont {J.~E.}\
  \bibnamefont {Kinsey}}, \bibinfo {author} {\bibfnamefont {M.}~\bibnamefont
  {Kotschenreuther}}, \bibinfo {author} {\bibfnamefont {A.~H.}\ \bibnamefont
  {Kritz}}, \bibinfo {author} {\bibfnamefont {L.~L.}\ \bibnamefont {Lao}},
  \bibinfo {author} {\bibfnamefont {J.}~\bibnamefont {Mandrekas}}, \bibinfo
  {author} {\bibfnamefont {W.~M.}\ \bibnamefont {Nevins}}, \bibinfo {author}
  {\bibfnamefont {S.~E.}\ \bibnamefont {Parker}}, \bibinfo {author}
  {\bibfnamefont {A.~J.}\ \bibnamefont {Redd}}, \bibinfo {author}
  {\bibfnamefont {D.~E.}\ \bibnamefont {Shumaker}}, \bibinfo {author}
  {\bibfnamefont {R.}~\bibnamefont {Sydora}}, \ and\ \bibinfo {author}
  {\bibfnamefont {J.}~\bibnamefont {Weiland}},\ }\href@noop {} {\bibfield
  {journal} {\bibinfo  {journal} {Phys. Plasmas}\ }\textbf {\bibinfo {volume}
  {7}},\ \bibinfo {pages} {969} (\bibinfo {year} {2000})}\BibitemShut {NoStop}%
\bibitem [{\citenamefont {G\"orler}\ \emph {et~al.}(2016)\citenamefont
  {G\"orler}, \citenamefont {Tronko}, \citenamefont {Hornsby}, \citenamefont
  {Bottino}, \citenamefont {Kleiber}, \citenamefont {Norscini}, \citenamefont
  {Grandgirard}, \citenamefont {Jenko},\ and\ \citenamefont
  {Sonnendrücker}}]{gorler2016}%
  \BibitemOpen
  \bibfield  {author} {\bibinfo {author} {\bibfnamefont {T.}~\bibnamefont
  {G\"orler}}, \bibinfo {author} {\bibfnamefont {N.}~\bibnamefont {Tronko}},
  \bibinfo {author} {\bibfnamefont {W.~A.}\ \bibnamefont {Hornsby}}, \bibinfo
  {author} {\bibfnamefont {A.}~\bibnamefont {Bottino}}, \bibinfo {author}
  {\bibfnamefont {R.}~\bibnamefont {Kleiber}}, \bibinfo {author} {\bibfnamefont
  {C.}~\bibnamefont {Norscini}}, \bibinfo {author} {\bibfnamefont
  {V.}~\bibnamefont {Grandgirard}}, \bibinfo {author} {\bibfnamefont
  {F.}~\bibnamefont {Jenko}}, \ and\ \bibinfo {author} {\bibfnamefont
  {E.}~\bibnamefont {Sonnendrücker}},\ }\href@noop {} {\bibfield  {journal}
  {\bibinfo  {journal} {Phys. Plasmas}\ }\textbf {\bibinfo {volume} {23}},\
  \bibinfo {pages} {072503} (\bibinfo {year} {2016})}\BibitemShut {NoStop}%
\bibitem [{\citenamefont {Ye}\ and\ \citenamefont {Chen}(2020)}]{ye2020}%
  \BibitemOpen
  \bibfield  {author} {\bibinfo {author} {\bibfnamefont {L.}~\bibnamefont
  {Ye}}\ and\ \bibinfo {author} {\bibfnamefont {Y.}~\bibnamefont {Chen}},\
  }\href {https://www.sciencedirect.com/science/article/pii/S0010465519303790}
  {\bibfield  {journal} {\bibinfo  {journal} {Comp. Phys. Comm.}\ }\textbf
  {\bibinfo {volume} {250}},\ \bibinfo {pages} {107050} (\bibinfo {year}
  {2020})}\BibitemShut {NoStop}%
\bibitem [{\citenamefont {Abiteboul}\ \emph {et~al.}(2011)\citenamefont
  {Abiteboul}, \citenamefont {Garbet}, \citenamefont {Grandgirard},
  \citenamefont {Allfrey}, \citenamefont {Ghendrih}, \citenamefont {Latu},
  \citenamefont {Sarazin},\ and\ \citenamefont {Strugarek}}]{abiteboul2011}%
  \BibitemOpen
  \bibfield  {author} {\bibinfo {author} {\bibfnamefont {J.}~\bibnamefont
  {Abiteboul}}, \bibinfo {author} {\bibfnamefont {X.}~\bibnamefont {Garbet}},
  \bibinfo {author} {\bibfnamefont {V.}~\bibnamefont {Grandgirard}}, \bibinfo
  {author} {\bibfnamefont {S.~J.}\ \bibnamefont {Allfrey}}, \bibinfo {author}
  {\bibfnamefont {P.}~\bibnamefont {Ghendrih}}, \bibinfo {author}
  {\bibfnamefont {G.}~\bibnamefont {Latu}}, \bibinfo {author} {\bibfnamefont
  {Y.}~\bibnamefont {Sarazin}}, \ and\ \bibinfo {author} {\bibfnamefont
  {A.}~\bibnamefont {Strugarek}},\ }\href@noop {} {\bibfield  {journal}
  {\bibinfo  {journal} {Phys. Plasmas}\ }\textbf {\bibinfo {volume} {18}},\
  \bibinfo {pages} {082503} (\bibinfo {year} {2011})}\BibitemShut {NoStop}%
\bibitem [{\citenamefont {G\"urcan}\ \emph {et~al.}(2007)\citenamefont
  {G\"urcan}, \citenamefont {Diamond}, \citenamefont {Hahm},\ and\
  \citenamefont {Singh}}]{gurcan2007}%
  \BibitemOpen
  \bibfield  {author} {\bibinfo {author} {\bibfnamefont {O.~D.}\ \bibnamefont
  {G\"urcan}}, \bibinfo {author} {\bibfnamefont {P.~H.}\ \bibnamefont
  {Diamond}}, \bibinfo {author} {\bibfnamefont {T.~S.}\ \bibnamefont {Hahm}}, \
  and\ \bibinfo {author} {\bibfnamefont {R.}~\bibnamefont {Singh}},\
  }\href@noop {} {\bibfield  {journal} {\bibinfo  {journal} {Phys. Plasmas}\
  }\textbf {\bibinfo {volume} {14}},\ \bibinfo {pages} {042306} (\bibinfo
  {year} {2007})}\BibitemShut {NoStop}%
\bibitem [{\citenamefont {G\"urcan}\ \emph {et~al.}(2010)\citenamefont
  {G\"urcan}, \citenamefont {Diamond}, \citenamefont {Hennequin}, \citenamefont
  {McDevitt}, \citenamefont {Garbet},\ and\ \citenamefont
  {Bourdelle}}]{gurcan2010}%
  \BibitemOpen
  \bibfield  {author} {\bibinfo {author} {\bibfnamefont {O.~D.}\ \bibnamefont
  {G\"urcan}}, \bibinfo {author} {\bibfnamefont {P.~H.}\ \bibnamefont
  {Diamond}}, \bibinfo {author} {\bibfnamefont {P.}~\bibnamefont {Hennequin}},
  \bibinfo {author} {\bibfnamefont {C.~J.}\ \bibnamefont {McDevitt}}, \bibinfo
  {author} {\bibfnamefont {X.}~\bibnamefont {Garbet}}, \ and\ \bibinfo {author}
  {\bibfnamefont {C.}~\bibnamefont {Bourdelle}},\ }\href@noop {} {\bibfield
  {journal} {\bibinfo  {journal} {Phys. Plasmas}\ }\textbf {\bibinfo {volume}
  {17}},\ \bibinfo {pages} {112309} (\bibinfo {year} {2010})}\BibitemShut
  {NoStop}%
\bibitem [{\citenamefont {Lu}\ \emph {et~al.}(2015)\citenamefont {Lu},
  \citenamefont {Wang}, \citenamefont {Diamond}, \citenamefont {Tynan},
  \citenamefont {Ethier}, \citenamefont {Chen}, \citenamefont {Gao},\ and\
  \citenamefont {Rice}}]{lu2015}%
  \BibitemOpen
  \bibfield  {author} {\bibinfo {author} {\bibfnamefont {Z.}~\bibnamefont
  {Lu}}, \bibinfo {author} {\bibfnamefont {W.~X.}\ \bibnamefont {Wang}},
  \bibinfo {author} {\bibfnamefont {P.~H.}\ \bibnamefont {Diamond}}, \bibinfo
  {author} {\bibfnamefont {G.}~\bibnamefont {Tynan}}, \bibinfo {author}
  {\bibfnamefont {S.}~\bibnamefont {Ethier}}, \bibinfo {author} {\bibfnamefont
  {J.}~\bibnamefont {Chen}}, \bibinfo {author} {\bibfnamefont {C.}~\bibnamefont
  {Gao}}, \ and\ \bibinfo {author} {\bibfnamefont {J.~E.}\ \bibnamefont
  {Rice}},\ }\href@noop {} {\bibfield  {journal} {\bibinfo  {journal} {Nucl.
  Fusion}\ }\textbf {\bibinfo {volume} {55}},\ \bibinfo {pages} {093012}
  (\bibinfo {year} {2015})}\BibitemShut {NoStop}%
\bibitem [{\citenamefont {Camenen}\ \emph {et~al.}(2009)\citenamefont
  {Camenen}, \citenamefont {Peeters}, \citenamefont {Angioni}, \citenamefont
  {Casson}, \citenamefont {Hornsby}, \citenamefont {Snodin},\ and\
  \citenamefont {Strintzi}}]{Camenen2009}%
  \BibitemOpen
  \bibfield  {author} {\bibinfo {author} {\bibfnamefont {Y.}~\bibnamefont
  {Camenen}}, \bibinfo {author} {\bibfnamefont {A.~G.}\ \bibnamefont
  {Peeters}}, \bibinfo {author} {\bibfnamefont {C.}~\bibnamefont {Angioni}},
  \bibinfo {author} {\bibfnamefont {F.~J.}\ \bibnamefont {Casson}}, \bibinfo
  {author} {\bibfnamefont {W.~A.}\ \bibnamefont {Hornsby}}, \bibinfo {author}
  {\bibfnamefont {A.~P.}\ \bibnamefont {Snodin}}, \ and\ \bibinfo {author}
  {\bibfnamefont {D.}~\bibnamefont {Strintzi}},\ }\href
  {https://link.aps.org/doi/10.1103/PhysRevLett.102.125001} {\bibfield
  {journal} {\bibinfo  {journal} {Phys. Rev. Lett.}\ }\textbf {\bibinfo
  {volume} {102}},\ \bibinfo {pages} {125001} (\bibinfo {year}
  {2009})}\BibitemShut {NoStop}%
\bibitem [{\citenamefont {Wang}\ \emph {et~al.}(2009)\citenamefont {Wang},
  \citenamefont {Hahm}, \citenamefont {Ethier}, \citenamefont {Rewoldt},
  \citenamefont {Lee}, \citenamefont {Tang}, \citenamefont {Kaye},\ and\
  \citenamefont {Diamond}}]{wang2009}%
  \BibitemOpen
  \bibfield  {author} {\bibinfo {author} {\bibfnamefont {W.~X.}\ \bibnamefont
  {Wang}}, \bibinfo {author} {\bibfnamefont {T.~S.}\ \bibnamefont {Hahm}},
  \bibinfo {author} {\bibfnamefont {S.}~\bibnamefont {Ethier}}, \bibinfo
  {author} {\bibfnamefont {G.}~\bibnamefont {Rewoldt}}, \bibinfo {author}
  {\bibfnamefont {W.~W.}\ \bibnamefont {Lee}}, \bibinfo {author} {\bibfnamefont
  {W.~M.}\ \bibnamefont {Tang}}, \bibinfo {author} {\bibfnamefont {S.~M.}\
  \bibnamefont {Kaye}}, \ and\ \bibinfo {author} {\bibfnamefont {P.~H.}\
  \bibnamefont {Diamond}},\ }\href
  {https://link.aps.org/doi/10.1103/PhysRevLett.102.035005} {\bibfield
  {journal} {\bibinfo  {journal} {Phys. Rev. Lett.}\ }\textbf {\bibinfo
  {volume} {102}},\ \bibinfo {pages} {035005} (\bibinfo {year}
  {2009})}\BibitemShut {NoStop}%
\bibitem [{\citenamefont {Camenen}\ \emph {et~al.}(2011)\citenamefont
  {Camenen}, \citenamefont {Idomura}, \citenamefont {Jolliet},\ and\
  \citenamefont {Peeters}}]{Camenen2011}%
  \BibitemOpen
  \bibfield  {author} {\bibinfo {author} {\bibfnamefont {Y.}~\bibnamefont
  {Camenen}}, \bibinfo {author} {\bibfnamefont {Y.}~\bibnamefont {Idomura}},
  \bibinfo {author} {\bibfnamefont {S.}~\bibnamefont {Jolliet}}, \ and\
  \bibinfo {author} {\bibfnamefont {A.~G.}\ \bibnamefont {Peeters}},\ }\href
  {\doibase 10.1088/0029-5515/51/7/073039} {\bibfield  {journal} {\bibinfo
  {journal} {Nucl. Fusion}\ }\textbf {\bibinfo {volume} {51}},\ \bibinfo
  {pages} {073039} (\bibinfo {year} {2011})}\BibitemShut {NoStop}%
\bibitem [{\citenamefont {Lu}(2015)}]{lu2015pop}%
  \BibitemOpen
  \bibfield  {author} {\bibinfo {author} {\bibfnamefont {Z.~X.}\ \bibnamefont
  {Lu}},\ }\href@noop {} {\bibfield  {journal} {\bibinfo  {journal} {Phys.
  Plasmas}\ }\textbf {\bibinfo {volume} {22}},\ \bibinfo {pages} {052118}
  (\bibinfo {year} {2015})}\BibitemShut {NoStop}%
\bibitem [{\citenamefont {Lu}\ \emph {et~al.}(2017)\citenamefont {Lu},
  \citenamefont {Fable}, \citenamefont {Hornsby}, \citenamefont {Angioni},
  \citenamefont {Bottino}, \citenamefont {Lauber},\ and\ \citenamefont
  {Zonca}}]{lu2017}%
  \BibitemOpen
  \bibfield  {author} {\bibinfo {author} {\bibfnamefont {Z.~X.}\ \bibnamefont
  {Lu}}, \bibinfo {author} {\bibfnamefont {E.}~\bibnamefont {Fable}}, \bibinfo
  {author} {\bibfnamefont {W.~A.}\ \bibnamefont {Hornsby}}, \bibinfo {author}
  {\bibfnamefont {C.}~\bibnamefont {Angioni}}, \bibinfo {author} {\bibfnamefont
  {A.}~\bibnamefont {Bottino}}, \bibinfo {author} {\bibfnamefont
  {P.}~\bibnamefont {Lauber}}, \ and\ \bibinfo {author} {\bibfnamefont
  {F.}~\bibnamefont {Zonca}},\ }\href@noop {} {\bibfield  {journal} {\bibinfo
  {journal} {Phys. Plasmas}\ }\textbf {\bibinfo {volume} {24}},\ \bibinfo
  {pages} {042502} (\bibinfo {year} {2017})}\BibitemShut {NoStop}%
\bibitem [{\citenamefont {Wang}\ and\ \citenamefont
  {Diamond}(2013)}]{wanglu2013}%
  \BibitemOpen
  \bibfield  {author} {\bibinfo {author} {\bibfnamefont {L.}~\bibnamefont
  {Wang}}\ and\ \bibinfo {author} {\bibfnamefont {P.~H.}\ \bibnamefont
  {Diamond}},\ }\href {https://link.aps.org/doi/10.1103/PhysRevLett.110.265006}
  {\bibfield  {journal} {\bibinfo  {journal} {Phys. Rev. Lett.}\ }\textbf
  {\bibinfo {volume} {110}},\ \bibinfo {pages} {265006} (\bibinfo {year}
  {2013})}\BibitemShut {NoStop}%
\bibitem [{\citenamefont {Hornsby}\ \emph {et~al.}(2018)\citenamefont
  {Hornsby}, \citenamefont {Angioni}, \citenamefont {Lu}, \citenamefont
  {Fable}, \citenamefont {Erofeev}, \citenamefont {McDermott}, \citenamefont
  {Medvedeva}, \citenamefont {Lebschy}, \citenamefont {Peeters} \emph
  {et~al.}}]{hornsby2018global}%
  \BibitemOpen
  \bibfield  {author} {\bibinfo {author} {\bibfnamefont {W.~A.}\ \bibnamefont
  {Hornsby}}, \bibinfo {author} {\bibfnamefont {C.}~\bibnamefont {Angioni}},
  \bibinfo {author} {\bibfnamefont {Z.~X.}\ \bibnamefont {Lu}}, \bibinfo
  {author} {\bibfnamefont {E.}~\bibnamefont {Fable}}, \bibinfo {author}
  {\bibfnamefont {I.}~\bibnamefont {Erofeev}}, \bibinfo {author} {\bibfnamefont
  {R.}~\bibnamefont {McDermott}}, \bibinfo {author} {\bibfnamefont
  {A.}~\bibnamefont {Medvedeva}}, \bibinfo {author} {\bibfnamefont
  {A.}~\bibnamefont {Lebschy}}, \bibinfo {author} {\bibfnamefont {A.~G.}\
  \bibnamefont {Peeters}},  \emph {et~al.},\ }\href@noop {} {\bibfield
  {journal} {\bibinfo  {journal} {Nucl. Fusion}\ }\textbf {\bibinfo {volume}
  {58}},\ \bibinfo {pages} {056008} (\bibinfo {year} {2018})}\BibitemShut
  {NoStop}%
\bibitem [{\citenamefont {Peeters}\ \emph {et~al.}(2011)\citenamefont
  {Peeters}, \citenamefont {Angioni}, \citenamefont {Bortolon}, \citenamefont
  {Camenen}, \citenamefont {Casson}, \citenamefont {Duval}, \citenamefont
  {Fiederspiel}, \citenamefont {Hornsby}, \citenamefont {Idomura},
  \citenamefont {Hein}, \citenamefont {Kluy}, \citenamefont {Mantica},
  \citenamefont {Parra}, \citenamefont {Snodin}, \citenamefont {Szepesi},
  \citenamefont {Strintzi}, \citenamefont {Tala}, \citenamefont {Tardini},
  \citenamefont {de~Vries},\ and\ \citenamefont {Weiland}}]{peeters_2011}%
  \BibitemOpen
  \bibfield  {author} {\bibinfo {author} {\bibfnamefont {A.~G.}\ \bibnamefont
  {Peeters}}, \bibinfo {author} {\bibfnamefont {C.}~\bibnamefont {Angioni}},
  \bibinfo {author} {\bibfnamefont {A.}~\bibnamefont {Bortolon}}, \bibinfo
  {author} {\bibfnamefont {Y.}~\bibnamefont {Camenen}}, \bibinfo {author}
  {\bibfnamefont {F.~J.}\ \bibnamefont {Casson}}, \bibinfo {author}
  {\bibfnamefont {B.}~\bibnamefont {Duval}}, \bibinfo {author} {\bibfnamefont
  {L.}~\bibnamefont {Fiederspiel}}, \bibinfo {author} {\bibfnamefont {W.~A.}\
  \bibnamefont {Hornsby}}, \bibinfo {author} {\bibfnamefont {Y.}~\bibnamefont
  {Idomura}}, \bibinfo {author} {\bibfnamefont {T.}~\bibnamefont {Hein}},
  \bibinfo {author} {\bibfnamefont {N.}~\bibnamefont {Kluy}}, \bibinfo {author}
  {\bibfnamefont {P.}~\bibnamefont {Mantica}}, \bibinfo {author} {\bibfnamefont
  {F.~I.}\ \bibnamefont {Parra}}, \bibinfo {author} {\bibfnamefont {A.~P.}\
  \bibnamefont {Snodin}}, \bibinfo {author} {\bibfnamefont {G.}~\bibnamefont
  {Szepesi}}, \bibinfo {author} {\bibfnamefont {D.}~\bibnamefont {Strintzi}},
  \bibinfo {author} {\bibfnamefont {T.}~\bibnamefont {Tala}}, \bibinfo {author}
  {\bibfnamefont {G.}~\bibnamefont {Tardini}}, \bibinfo {author} {\bibfnamefont
  {P.}~\bibnamefont {de~Vries}}, \ and\ \bibinfo {author} {\bibfnamefont
  {J.}~\bibnamefont {Weiland}},\ }\href@noop {} {\bibfield  {journal} {\bibinfo
   {journal} {Nucl. Fusion}\ }\textbf {\bibinfo {volume} {51}},\ \bibinfo
  {pages} {094027} (\bibinfo {year} {2011})}\BibitemShut {NoStop}%
\bibitem [{\citenamefont {Diamond}\ \emph {et~al.}(2005)\citenamefont
  {Diamond}, \citenamefont {Itoh}, \citenamefont {Itoh},\ and\ \citenamefont
  {Hahm}}]{Diamond_2005}%
  \BibitemOpen
  \bibfield  {author} {\bibinfo {author} {\bibfnamefont {P.~H.}\ \bibnamefont
  {Diamond}}, \bibinfo {author} {\bibfnamefont {S.-I.}\ \bibnamefont {Itoh}},
  \bibinfo {author} {\bibfnamefont {K.}~\bibnamefont {Itoh}}, \ and\ \bibinfo
  {author} {\bibfnamefont {T.~S.}\ \bibnamefont {Hahm}},\ }\href@noop {}
  {\bibfield  {journal} {\bibinfo  {journal} {Plasma Phys. Controlled Fusion}\
  }\textbf {\bibinfo {volume} {47}},\ \bibinfo {pages} {R35} (\bibinfo {year}
  {2005})}\BibitemShut {NoStop}%
\bibitem [{\citenamefont {Rice}\ \emph {et~al.}(2011)\citenamefont {Rice},
  \citenamefont {Duval}, \citenamefont {Reinke}, \citenamefont {Podpaly},
  \citenamefont {Bortolon}, \citenamefont {Churchill}, \citenamefont
  {Cziegler}, \citenamefont {Diamond}, \citenamefont {Dominguez}, \citenamefont
  {Ennever}, \citenamefont {Fiore}, \citenamefont {Granetz}, \citenamefont
  {Greenwald}, \citenamefont {Hubbard}, \citenamefont {Hughes}, \citenamefont
  {Irby}, \citenamefont {Ma}, \citenamefont {Marmar}, \citenamefont
  {McDermott}, \citenamefont {Porkolab}, \citenamefont {Tsujii},\ and\
  \citenamefont {Wolfe}}]{rice2011}%
  \BibitemOpen
  \bibfield  {author} {\bibinfo {author} {\bibfnamefont {J.~E.}\ \bibnamefont
  {Rice}}, \bibinfo {author} {\bibfnamefont {B.~P.}\ \bibnamefont {Duval}},
  \bibinfo {author} {\bibfnamefont {M.~L.}\ \bibnamefont {Reinke}}, \bibinfo
  {author} {\bibfnamefont {Y.~A.}\ \bibnamefont {Podpaly}}, \bibinfo {author}
  {\bibfnamefont {A.}~\bibnamefont {Bortolon}}, \bibinfo {author}
  {\bibfnamefont {R.~M.}\ \bibnamefont {Churchill}}, \bibinfo {author}
  {\bibfnamefont {I.}~\bibnamefont {Cziegler}}, \bibinfo {author}
  {\bibfnamefont {P.~H.}\ \bibnamefont {Diamond}}, \bibinfo {author}
  {\bibfnamefont {A.}~\bibnamefont {Dominguez}}, \bibinfo {author}
  {\bibfnamefont {P.~C.}\ \bibnamefont {Ennever}}, \bibinfo {author}
  {\bibfnamefont {C.~L.}\ \bibnamefont {Fiore}}, \bibinfo {author}
  {\bibfnamefont {R.~S.}\ \bibnamefont {Granetz}}, \bibinfo {author}
  {\bibfnamefont {M.~J.}\ \bibnamefont {Greenwald}}, \bibinfo {author}
  {\bibfnamefont {A.~E.}\ \bibnamefont {Hubbard}}, \bibinfo {author}
  {\bibfnamefont {J.~W.}\ \bibnamefont {Hughes}}, \bibinfo {author}
  {\bibfnamefont {J.~H.}\ \bibnamefont {Irby}}, \bibinfo {author}
  {\bibfnamefont {Y.}~\bibnamefont {Ma}}, \bibinfo {author} {\bibfnamefont
  {E.~S.}\ \bibnamefont {Marmar}}, \bibinfo {author} {\bibfnamefont {R.~M.}\
  \bibnamefont {McDermott}}, \bibinfo {author} {\bibfnamefont {M.}~\bibnamefont
  {Porkolab}}, \bibinfo {author} {\bibfnamefont {N.}~\bibnamefont {Tsujii}}, \
  and\ \bibinfo {author} {\bibfnamefont {S.~M.}\ \bibnamefont {Wolfe}},\ }\href
  {https://doi.org/10.1088/0029-5515/51/8/083005} {\bibfield  {journal}
  {\bibinfo  {journal} {Nucl. Fusion}\ }\textbf {\bibinfo {volume} {51}},\
  \bibinfo {pages} {083005} (\bibinfo {year} {2011})}\BibitemShut {NoStop}%
\end{thebibliography}
\end{document}